\documentclass[manuscript]{acmart}

\AtBeginDocument{%
  \providecommand\BibTeX{{%
    \normalfont B\kern-0.5em{\scshape i\kern-0.25em b}\kern-0.8em\TeX}}}

\copyrightyear{2022} 
\acmYear{2022} 
\setcopyright{acmlicensed}
\acmConference[TEI '22]{Sixteenth International Conference on Tangible, Embedded, and Embodied Interaction}{February 13--16, 2022}{Daejeon, Republic of Korea}
\acmBooktitle{Sixteenth International Conference on Tangible, Embedded, and Embodied Interaction (TEI '22), February 13--16, 2022, Daejeon, Republic of Korea}
\acmPrice{15.00}
\acmDOI{10.1145/3490149.3501331}
\acmISBN{978-1-4503-9147-4/22/02}

\usepackage{xspace}
\usepackage{calc}

\newcommand{\ef}{{Ephemeral Fabrication}\xspace}
\newcommand{\tf}{Temporary Fabrication\xspace}
\newcommand{\llf}{Long-Lasting Fabrication\xspace}
\newcommand{\pf}{Permanent Fabrication\xspace}
\newcommand{\system}{Draupnir\xspace}

\begin{document}

\title[Ephemeral Fabrication]{Ephemeral Fabrication: \\Exploring a Ubiquitous Fabrication Scenario of Low-Effort, In-Situ Creation of Short-Lived Physical Artifacts}
\author{Evgeny Stemasov}
\email{evgeny.stemasov@uni-ulm.de}
\orcid{0000-0002-3748-6441}
\author{Alexander Botner}
\email{alexander.botner@kopsi.dev}
\author{Enrico Rukzio}
\email{enrico.rukzio@uni-ulm.de}
\affiliation{%
  \institution{Institute of Media Informatics, Ulm University}
  \city{Ulm}
  \country{Germany}}

\author{Jan Gugenheimer}
\orcid{0000-0002-6466-3845}
\affiliation{%
  \institution{Télécom Paris -- LTCI, Institut Polytechnique de Paris}
  \city{Paris}
  \country{France}}
\email{jan.gugenheimer@telecom-paris.fr}

\renewcommand{\shortauthors}{Stemasov et al.}

\begin{abstract}
  Personal fabrication empowers users to create objects increasingly easier and faster.
This continuous decrease in effort evokes a speculative scenario of Ephemeral Fabrication (EF), enabled and amplified by emerging paradigms of mobile, wearable, or even body-integrated fabrication.
EF yields fast, temporary, in-situ solutions for everyday problems (e.g., creating a protective skin, affixing a phone).
Users solely create those, since the required effort is negligible.
We present and critically reflect on the EF scenario, by exploring current trends in research and building a body-worn fabrication device. 
EF is a plausible extrapolation of current developments, entailing both positive (e.g., accessibility) and negative implications (e.g., unsustainability).
Using speculative design methodology to question the trajectory of personal fabrication, we argue that to avert the aftermath of such futures, topics like sustainability can not remain an afterthought, but rather be situated in interactions themselves: through embedded constraints, conscious material choice, and constructive embedding of ephemerality.
\end{abstract}





\maketitle

\section{Introduction}
\label{sec:introduction}
    \begin{figure}[t!]
        \centering
        \includegraphics[width=\minof{\columnwidth}{0.8\textwidth}]{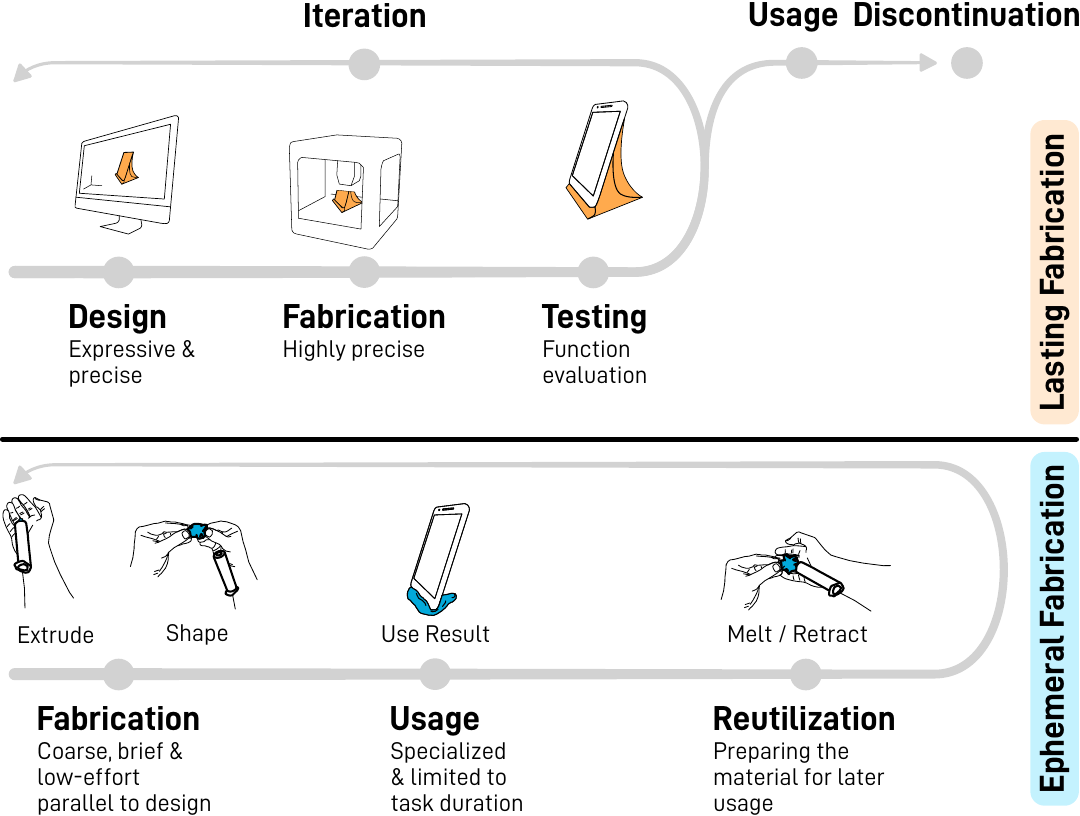}
        \caption{
        Conceptual flows of more ''traditional'' personal fabrication (top) in contrast to \ef (bottom). 
        Currently, design iterations for physical artifacts often involve a positive net material use to achieve a \textit{lasting} result. 
        If users are handed an effortless way to fabricate, resulting artifacts may transition to be non-lasting and explicitly designed to be quick to create for a short-lived use-case. Such an ephemeral usage dictates embedding reuse by design over the emphasis of (design) iterations.
        }~\label{fig:teasernew}
    \end{figure}

    Advances in personal fabrication technologies enable users to create artifacts and tools specifically tailored for their requirements~\cite{baudischPersonalFabrication2017,chenRepriseDesignTool2016,roumenMobileFabrication2016}.
    CAD software and devices like 3D-printers provide outstanding precision, currently at the cost of arguably slow design and fabrication processes~\cite{muellerFaBrickationFast3D2014}.
    With progress in research and practice, these processes are becoming both simpler~\cite{yungPrinty3DInsituTangible2018,savageMakersMarksPhysical2015} and faster~\cite{muellerWirePrint3DPrinted2014} for users of any skill level.
    For the creation of \emph{lasting} artifacts and adaptations, investing time (to learn, design, fabricate, and iterate) is an appropriate tradeoff. 
    This tradeoff, along with a fairly low permeation of personal fabrication through society, likewise currently hinders the use of technologies like 3D-printing in short-lived contexts.
    When we look ahead, even lower effort and time expenditure are likely to manifest as a result of current research efforts (increased ease-of-use, mobility, effortlessness).
    The shorter the design and fabrication process, the more spontaneous its use may be and the more scenarios it can be employed in, ultimately making personal fabrication an ubiquitous procedure and concept~\cite{stemasovRoadUbiquitousPersonal2021}. 
	However, our current stance on sustainability and the way it is considered for novel systems for personal fabrication is unfit for a future of widespread, ubiquitous personal fabrication: it is often deferred to recycling processes, material science or not considered at all, as novel systems aim to create \emph{lasting} artifacts.
	We argue that systems that embed sustainable material use in their interactions themselves (e.g., \emph{Unfabricate} \cite{wuUnfabricateDesigningSmart2020}) are indispensable for further research in personal fabrication, if the path of ever simpler and faster personal fabrication is to be followed.
	In particular, as the associated processes (e.g., design/modeling, fabrication) become easier, faster, and more effortless.
	At this point, this future progression creates a Collingridge dilemma\footnote{also known as Collingridge’s dilemma of control}~\cite{brockmanThisExplainsEverything2013,collingridgeSocialControlTechnology1982}:
	Before personal fabrication has permeated society, it can still be regulated and influenced by research and practice.
	The future of widespread adoption of personal fabrication, however, is hard to predict and explore.
	When personal fabrication has become widespread already, change, for instance concerning responsible and sustainable use, is far harder to achieve.
	To engage with this dilemma, we employ a speculative design approach combined with physical prototyping.\\

	In this work, we explore, present, and criticize the \ef scenario, which we argue is a direction the current field of research on personal fabrication is slowly gravitating towards (i.e., faster and easier access to fabrication technology for non-experts).
    It emerges when fabrication becomes an easy and near-effortlessly accessible human ability. 
    We arrived at this discussion after exploring the vision of personal fabrication becoming a widespread (mechanical) augmentation of human users.
    This culmination is relevant to consider as a speculative and \emph{possible}~\cite{mitrovicIntroductionSpeculativeDesign2015} future of (personal) fabrication, as it shifts the importance of sustainable material use, (numerical) precision, and the value of artifacts, in contrast to today's perspectives.
    Unlike today's notions of personal fabrication, this speculative future may not be focused on long-lasting artifacts, precision, or long design procedures. 
    Rather, it may additionally spread to coarse, short-term solutions to potentially mundane, everyday problems.
    While not the only future use of (personal) fabrication, \ef may become a dominant one, with considerable influence on the amount of waste generated.
    This more regularly and heavily employed type of personal fabrication then underlies other constraints and exhibits novel requirements: industry-grade precision, processes, and appearance may drop in terms of importance while sustainable use of material becomes indispensable (figure \ref{fig:teasernew}), as the goal is not a highly precise lasting artifact, but rather one that can quickly address a temporary challenge.
    We explored and formulated the scenario of \ef through a diegetic prototype~\cite{tanenbaumDesignFictionalInteractions2014,boschSciFiWriterBruce2012}, \system\footnote{Draupnir - ''the dripper''~\cite{orchardCassellDictionaryNorse1998}. Named after a ring Odin's from Norse mythology, which replicates itself: \emph{''From it do eight of like weight fall on every ninth night''}~\cite{bellowsPoeticEdda1936}. While physically impossible, it applies generation of matter to a body-worn (i.e., wearable) artifact.}, which may provide users raw material for near-effortless, tangible fabrication activities for everyday mechanical tasks.
    It embodies a future of ubiquitous, quick access to fabrication devices and low-effort processes -- which may entail positive consequences, like augmented human problem-solving, but also negative ones, like its impact on sustainability and material use.
    In this work, we present, explore and reflect on the scenario of \ef by employing speculative design methodology.
    We explore this scenario further by developing and implementing \system as a \emph{functional} design artifact with today's means and materials: a body-worn add-on to users' hands which provides the ability to extrude malleable material (Polycaprolactone, PCL) controlled by an EMG-sensor (Electromyography), keeping it malleable and providing it right to the users' fingertips to be shaped. 
    This allows potential users to adapt their environment near-instantly by fabricating small-scale artifacts, for instance, to briefly prop up a phone.
    By relying on a limited amount of reusable material, \system embeds and enforces a degree of re-use, instead of deferring it to subsequent processes (such as slow, eventual biodegradation).\\

    With this work, we want to present and critically reflect upon one potential ''fabrication scenario'', \ef, that could potentially arise within a future of ubiquitous personal fabrication.
    Through the lens of this fabrication scenario, we aim to add to the discussion on effort, time, and sustainability in personal fabrication through the following contributions:
    \begin{itemize}
        \item \textbf{Definition and categorization of \ef} in a theoretical framework focusing on artifact longevity and effort as core dimensions. \ef is a fabrication \emph{scenario} involving low-effort, in-situ artifacts for highly specialized and short-lived problems. It describes a future where low-effort fabrication is widespread and requires deep embedding of sustainability in novel systems.
        \item \textbf{Reflection and discussion of sustainability} within the context of personal fabrication through the lens of current literature and a diegetic prototype, \system.
        \item Presentation of a set of \textbf{usage scenarios and interaction patterns} arising from and relevant for the \ef scenario. We additionally report on the implementation of \system, a functional design artifact to explore the \ef interaction scenario, along with practical insights into the development of this device for \ef. 
    \end{itemize}
    
    Finally, we want to emphasize that the goal of this work is not to promote either \ef or \system as \textit{''solutions''}. 
    We see \ef as an exaggerated but realistic future scenario following the current trends in personal fabrication. 
    Additionally, we see \system both as a diegetic and a physical prototype, to be mainly a tool to explore the \ef scenario, while openly reporting on its development and influence. 
    A future system could take any other shape or form other than \system (e.g., through mobile tools, like hot glue guns, or mobile material stockpiles, like users carrying clay with them), and we are not arguing that \system is the \textit{''right''} design to enable \ef.
    However, designing, building, and interacting with \system helped us further understand and explore \ef. 
    Ultimately, this work wants to discuss current trends in the field of personal fabrication using a design methodology and engage in a discussion around sustainable personal fabrication.

\section{Design Process}
\label{sec:designprocess}
    The following paragraphs describe the interconnected process that influenced the concept of \ef (EF) and the development of \system, both as a diegetic and an actual prototype, along with the scenarios -- all of which are detailed in subsequent sections of this work.
    We consider our approach to follow speculative design practices \cite{dunneSpeculativeEverythingDesign2013,augerSpeculativeDesignCrafting2013}.
    Our premises for speculation consisted of 3 aspects: 1) increasingly easier and faster personal fabrication coupled with their increasing adoption and domestication, 2) nature-inspired and technology-mediated human augmentation, and 3) historical developments in computing, expanding the applicability of previously expert-only tools to mundane tasks. 
    This yielded our depiction of a \emph{plausible} future surrounding our diegetic prototype of Draupnir (section \ref{subsec:narrative}), along with a formalization of Ephemeral Fabrication as a possible future fabrication scenario.
    This future may or may not be a \emph{desirable} one.
    To further explore this future, we engaged in material explorations, ultimately building an early, functional version of Draupnir with current means (section \ref{sec:prototype}).

    \subsection{A Day in the Life of a Fabrication-augmented User}
        \label{subsec:narrative}
        
        \begin{figure*}[ht]
            \includegraphics[width=\linewidth]{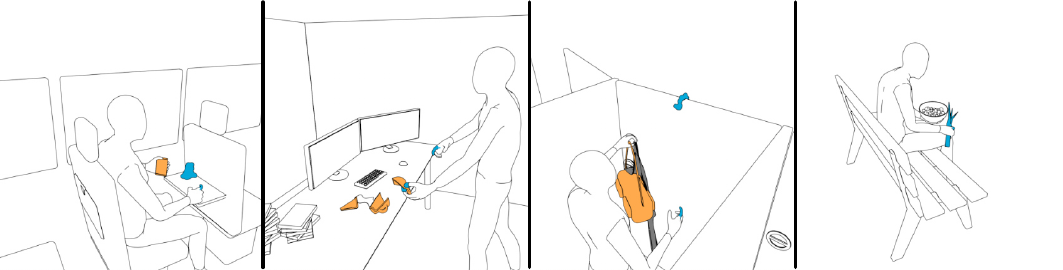}
            \caption{A day in the life of a future fabrication-augmented user. They are able to solve mundane everyday challenges, as fabrication became a low-effort ability for them, not a tool only to be used for complex challenges. From left to right: creating a phone mount to use during a commute, quickly creating thimbles to pick up shards, creating a hook to suspend one's belongings if none is available, creating a fork to consume a meal.}~\label{fig:narrative}
        \end{figure*}
        
        With this paragraph, we want to construct and present a narrative of a future where anyone may possess an augmentation giving them access to personal fabrication.
        While this may already be the case today (e.g., in well-equipped workshops or crafty individuals carrying tools and resources with them), we want to focus on extremely low access times and extremely low effort needed.
        This progression (i.e., faster~\cite{muellerWirePrint3DPrinted2014}, more mobile~\cite{roumenMobileFabrication2016,peekPopfabCasePortable2017}, easier to learn~\cite{bermanAnyoneCanPrint2020,bermanHowDIYMetaDesignTools2021}) makes the use of personal fabrication relevant for far more scenarios than now.
        In the following scenarios, \system is treated and presented as a diegetic prototype: a material-deploying implant in one's finger, naturally controlled by a user, enabling constant use of it and the creation of ephemeral artifacts.
        We consider this constellation to be a \emph{possible} expression of \ef in the future.
        However, \ef may likewise be enabled through other means or ''systems'', such as mobile 3D-printing pens, or even stationary fabrication devices.
        It is conceivable that fast stationary devices can be used to create ephemeral artifacts.
        Likewise, it is possible to generate lasting artifacts with devices such as 3D-printing pens.
        However, low access and ''design'' times (i.e., ideally none) as enabled by a tool like \system lend themselves to be applied for \ef.\\
        
        \newcommand{\username}[0]{Max\xspace}
        \newcommand{\material}[0]{PCLx\xspace}
        
        \emph{
        ''Imagine a future where control over matter and shape is right at your fingertips.
        No thinking, no designing, just coarse-yet-functional solutions for everyday needs.'' -- that was the pitch for the first versions of \system.
        \username was one of the early adopters, eager to try the device, without really knowing what it may be needed for. 
        It wasn't as easy as advertised, but \username, over time, found out where \system actually shines: improvisation.
        By now, it has become a commodity, just as 3D-printers became one in the 30s.
        The advertisements now focus on how \system augments and alters cognition: it's yet another ability in the repertoire of the modern person.
        \system has indeed altered the way users interact with their physical spaces: users are now in control of matter, for better or for worse.
        The absence of a tool or an object for a task does not mean that this task has to be delayed anymore.
        Users now can fabricate whatever they need, whenever they need it. 
        Discomfort while grabbing virtually anything can now be ''resolved'' with a quick blob of material from \system.\newline\newline
        \username recently received an upgrade for their \system device.
        It's mainly a better version of the \material material, now suited for more re-use cycles.
        The older one degraded after a few uses, drawing the attention of environment activists.
        After all, users of \system kept buying material cartridges, instead of re-using the material from objects they made.
        \username is now on their way to work; a bumpy ride in public transport.
        They would like to continue watching the series they started yesterday, but holding the screen gets tiring and it just doesn't want to stay upright at the table.
        \username quickly drags a thin slice of \material out of the upgrade in their right hand and, in a rapid  motion, fashions an L-shaped structure able to hold the screen at a comfortable viewing angle.
        When it's their turn to disembark, they retract the screen-mount back into the \system system.
        \username is now at work, toiling in front of yet another screen.
        A mug was dropped at their table in the office and shattered.
        ''I could just go and search for safety gloves'', \username thinks ''... or a dustpan''.
        They don't, as it's much faster to just fabricate two small thimbles, one for the thumb and one for the index finger, grab the shards and drop them into the trash.
        The smaller ceramic pieces stick to the still malleable \material material, rendering it unfit for re-use later.
        On the way home, \username stops by a public restroom.
        There's no clean spot for their coat and bag, nor a coat hook at the door to use.
        A minor inconvenience.
        \username briefly extrudes some material from \system, sculpting it into an s-shape to hang over the top of the door.
        After the material solidifies, they can hang their belongings from it.
        Afterwards, they again retract the material back into \system.
        Before coming home, \username wants to grab a meal to eat somewhere in the park and buys a prepared salad from a convenience store.
        Single-use cutlery was considered the bane upon the environment once. 
        It's still a part of everyday life, just as many other single-use objects, like packaging.
        \username forgot to take such a fork from the checkout at the store, noticing this oversight only when they've already found a spot in the park to eat at.
        They quickly fabricate something looking like a trident, irregular, ugly even, but they are able to eat the salad they bought.\newline\newline
        \system v23.0 is already announced, touting revolutionary improvements by using ''perfect red''\footnote{c.f. Ishii et al. ''Radical Atoms''~\cite{ishiiRadicalAtomsTangible2012}} as a material.
        ''Do I really need this?'' \username ponders, ''... I'm not an engineer after all''.}\\
        
        When comparing these tasks with ''Mobile Fabrication'' by Roumen et al. \cite{roumenMobileFabrication2016}, they may appear similar at first glance.
        However, these use-cases of \ef may exhibit no immediate need, no necessity to be enacted, and certainly solve no revolutionary engineering challenge.
        They do not necessarily emerge from negative consequences that may arise if the user abstains from fabricating the artifacts (e.g., dangers in traffic if one's bike lamp is not fixed \cite{roumenMobileFabrication2016}).
        Rather, they emerge from various minuscule benefits, scattered across a user's daily life.
        However, if the effort becomes negligibly low (e.g, through a highly accessible and fast process), they become an everyday act, making the consideration of reuse and sustainability a necessity a priori.
        We consider the actual tool to be used to engage in \ef to be of secondary importance, but highlight certain aspects that may enable it in section \ref{subsec:eftable}.
    
    \subsection{Why \ef is a Possible and Plausible Future}
        Our work started out by observing the current progress of miniaturization~\cite{roumenMobileFabrication2016}, augmentation~\cite{peng3DPrinterInteractive2016} and acceleration~\cite{muellerWirePrint3DPrinted2014} of personal fabrication tools and a future extrapolating of this towards the extreme of having constant access to fast personal fabrication, similarly to examples from nature (e.g., spiders or bees). 
        Therefore, the starting point for this work was the premise of imagining the speculative future of (personal) fabrication as an inherent human ability.
        We consider this to be one of many possible expressions of ubiquitous (personal) fabrication~\cite{gershenfeldDesigningRealityHow2017,stemasovRoadUbiquitousPersonal2021} in the future.
        Fabrication can be leveraged through tool use and available materials (e.g., birds fabricating nests), but also through an integrated, always-available and always-feasible ability (e.g., spiders fabricating webs). 
        The latter is the fictional space, where we treat fabrication as an immediate personal human augmentation.
        A viable predecessor to body-\emph{integrated} fabrication interfaces, is the concept of body-\emph{worn} or wearable personal fabrication~\cite{fernandesHumanAugmentationWearables2016}, which we evaluated initially as a step beyond mobile fabrication~\cite{roumenMobileFabrication2016} by building \system as an actual prototype with the means available to us right now.
        
        \begin{figure}[ht!]
            \centering
            \includegraphics[width=\minof{\columnwidth}{0.85\textwidth}]{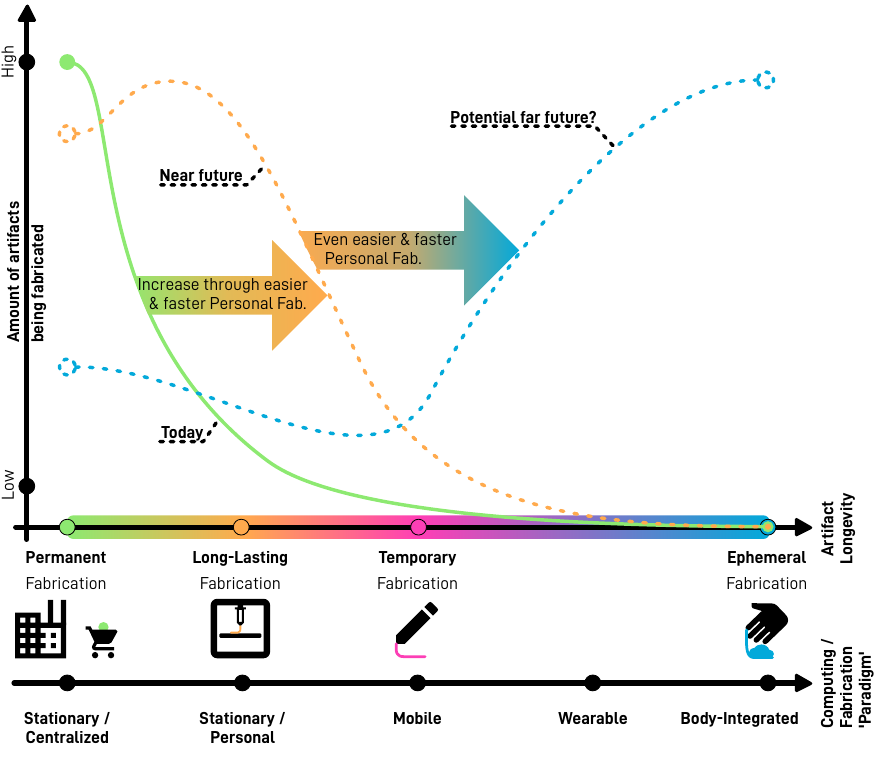}
            \caption{Research in personal fabrication has already enabled a shift from mainframe-like fabrication to personal fabrication, similarly to computing (bottom x-axis).
            This can be likened to the development of ubiquitous computing~\protect\cite{weiserComputer21St1991}.
            Following these computing/fabrication paradigms, a temporal dimension with respect to fabrication and use duration can be approximated (top x-axis), ranging from (industry-grade) permanent fabrication to \ef, which we consider the culmination of low effort and access times to personal fabrication processes. The used paradigm indicates a likely artifact longevity, but may not be restricted to one (e.g., one may use a personal stationary printer to create ephemeral artifacts, if it is fast enough). A near future with more personal fabrication processes is already manifesting. A potential far future may consist of an overwhelming amount of ephemeral artifacts, if time and effort are constantly reduced.}~\label{fig:extrapolation}
        \end{figure}
    
        Parallels can be drawn to current and future computing paradigms outlined by Mark Weiser~\cite{weiserComputer21St1991,weiserComingAgeCalm1997} (figure \ref{fig:extrapolation}): just as a stationary home computer, for instance, allows users to author text at home, a stationary home fabrication device (e.g., a hobbyist 3D-printer) can fabricate intricate artifacts, with time and effort required for the design and the fabrication process itself.
        A portable or mobile device, like a smartphone enables users to conduct some of the tasks, previously done with the home computer, in a mobile or nomadic setting. 
        In this progression to more ubiquitous usage, tradeoffs have to be made - text input speeds on smartphones can be slower and wielding a 3D-printing pen often incurs a loss of precision, as we replace NC (numerical control) with manual operation. 
        Mobile Fabrication \cite{roumenMobileFabrication2016} may refer to the use of a 3D-printing pen, a mobile 3D-printer, or material like Sugru\footnote{\url{sugru.com}, Accessed 13.7.2021}, which users may carry around in their pockets or bags.
        This reduces access times further, making fabrication potentially more feasible for more tasks.
        The inhibition threshold to apply personal fabrication to a given challenge is lowered.
        Wearable technology reduces access times in a similar fashion, while also constricting the input and output expressivity further.
        Texts authored on a smartwatch are often merely quick responses to messages.
        Likewise, artifacts made with a wearable fabrication device would likely be coarse, but may still fulfill their intended function and be quick to create.
        In our design approach, we followed this line of thought, drawing parallels to personal computing and tried to deduct systems and concepts that would exist in such a future.
        Similarly, such systems would be the parallel to wearable computing research.
        These parallels between computing and fabrication technology can also be (coarsely) mapped to the artifacts they may produce and their longevity (figure \ref{fig:extrapolation}, bottom x axis). 
        With reduced access and fabrication times, the artifact that results may be applied to brief and mundane tasks, which may do without industry-grade precision or visual appeal.
        By being highly context-specific, it may not make sense to re-use the result (as done with long-lasting artifacts), thereby making the re-use of the material mandatory.
        With increasingly lower effort and time requirements being made possible by researchers and practitioners, a growing subset of artifacts being made may become \emph{ephemeral}.
        This can be likened to today's pervasiveness of single-use plastics and notions of a ''throw-away society'' embracing such mass-produced single-use artifacts.
        \system was a result of this design exercise which embodies personal fabrication as a human ability.
        We initially conceived it as a wearable system, with respect to the near future, and as a body-integrated system, with respect to a more distant future following similar ideas with respect to computing paradigms itself.
        As \ef is a brief activity, it may emerge to be applied to overwhelming numbers of everyday small challenges, making the impact of even a few users an immense one (figure \ref{fig:extrapolation}).
        
    \subsection{How Ephemeral Fabrication May Manifest}
        To further immerse our design process into the created fictional future of \ef through wearable or body-\textit{integrated} fabrication, we created a working prototype based on the idea (i.e., diegetic prototype) of \system. 
        The implementation of \system is a body-\textit{worn} (wearable) paste-extrusion device, controlled though EMG.
        It dispenses up to 20 grams of malleable Polycaprolactone (PCL), which, when solidified, can be reliably used for small mechanical tasks.
        Furthermore, PCL can also be re-heated for later reuse.
        The creation and later usage of the prototype helped us to explore this speculative future in more depth and was not done with the goal of having a finished product, but was done as a vehicle to engage with the fictional scenario of having always-available fabrication abilities and opportunities. 
        This immersion with this fictional scenario helped us discover, define and understand the arising application scenario of \ef, which we further categorized with existing literature.
        The engineering procedure initially involved the exploration of materials, where attempts were made to evaluate suitable substances that may enable \ef. 
        The initial requirements were reuseablity on one hand, but also suitability for basic mechanical tasks (e.g., affixing or mounting objects).
        As mentioned earlier, our incarnation of \ef initially aimed to generate and embody a \emph{positive} future outcome, which certainly requires reuseability as an inherent component of the system.
        These decisions for a positive future made us also reflect on potentially dystopian scenarios where sustainable material use is not considered at a point where personal fabrication became an ubiquitous technology.
        Additional requirements were the ability to shape the material without additional tools (i.e., malleability and skin-safety).
        This allowed for initial explorations of feasibility with materials like wax, PLA (Polylactic Acid) or PCL.
        
    \subsection{How Draupnir Shaped \ef}
        Consecutively, the prototype was then used to brainstorm and explore potential application scenarios of constant, ideally nearly effortless access to a personal fabrication tool. 
        This resulted in the realization that the moment one has the access to fast and effortless personal fabrication, application scenarios are extended from precise and static artifacts towards short lived (\textbf{ephemeral}) solutions to everyday problems that can dynamically change (e.g., the mounting position of a phone).
        A ''traditionally'' fabricated solution -- one that often requires the user to design, fabricate and often iterate (c.f., fig. \ref{fig:teasernew}) -- is not feasible for such tasks, taking both time and effort.
        Such mundane problems also include areas where users may simply improvise with materials at hand (e.g., using a mug to prop up a phone), if such materials happen to be available.
        This availability is not guaranteed, especially in mobile contexts.
        In cases where the material is not available in the environment, an \ef device may provide this resource for ephemeral problem-solving.
        The vision of \ef was developed further by treating \system as both a diegetic and an actual prototype.
        This yielded the more specific interaction patterns, where inherent disadvantages of a quick and coarse fabrication process are compensated for.
        The implemented version of Draupnir uses digital components (i.e., to keep the material ready and malleable, or to control extrusion), but does not yield digital \emph{precision}, in favor of brief and coarse manual fabrication activities.
        We consider Draupnir to be a \emph{personal fabrication device} nonetheless, as it is a highly personal and pervasive interface enabling fabrication.
        Precision may re-emerge in further iterations of it, for instance through the use of smarter materials (e.g., ''perfect red''~\cite{ishiiRadicalAtomsTangible2012}) or motion guidance (e.g., through use of electrical muscle stimulation~\cite{lopesMuscleplotterInteractiveSystem2016}).\\
    
    With this work, we aimed to explore what \emph{might be} and question what \emph{ought to be} \cite{zimmermanResearchDesignMethod2007}.
    Accelerating personal fabrication and reducing the effort needed for it, is a promising direction for interaction design and engineering alike. 
    This process is already set in motion, and yields various intriguing research contributions.
    What \emph{might be}, is that personal fabrication is quick, effortless and unsustainable, with artifacts losing any value they gain through a process conducted by users.
    What \emph{ought to be}, is that personal fabrication should aim for re-use and sustainability by design. 
    Exploring this speculative future of \ef highlights the potential dangers of quick, effortless, and ubiquitous (personal) fabrication.
    Drawing a parallel to digital content creation and remixing such as videos and images, we can see that the rise of simple tools for creating new digital content (e.g., TikTok for videos and Instagram for images) reduces the average personal value of each artifact but also leads to an unprecedented amount of content and content creators~\cite{stemasovRoadUbiquitousPersonal2021}. 
    Enabling the same simple form of creation for personal fabrication -- which is a more distant future -- comes with more direct implications for our environment and our consumption of energy and materials. 

\section{Ephemeral Fabrication}
    \label{sec:ephemeralfab}
    In the following, we will present a grounding of \ef in current literature and outline its differences to prior concepts such as mobile fabrication.  
    The term ''ephemeral'' is defined as ''lasting a very short time''~\cite{merriam-websterincorporatedDefinitionEphemeral2019} or ''lasting one day only''~\cite{merriam-websterincorporatedDefinitionEphemeral2019}.
    In contrast to that, personal -- and particularly digital -- fabrication often aims for longer-lasting artifacts and replacing centralized, industrial processes with more decentralized ones~\cite{gershenfeldDesigningRealityHow2017,baudischPersonalFabrication2017}, enabling a high degree of personalization (c.f. figure \ref{fig:extrapolation}).
    To successfully create and fabricate such artifacts, tools are needed: users generally have to specify their design in CAD/CAM software, transfer it to the fabrication machine of their choice, and, optionally, post-process it (e.g., through the removal of support structures, painting or sanding).
    All steps, including the previously required learning of the process, require some degree of \emph{effort}~\cite{stemasovRoadUbiquitousPersonal2021}.
    Effort is antithetical to the concept of ephemeral artifacts.
    Why bother with hours of modeling, printing/milling/cutting, iterating, if the result is going to be discarded after a few minutes or hours of use?
    Unless the task is of the highest importance and urgency~\cite{roumenMobileFabrication2016}, the inhibition to undertake such a lengthy process is far too high.
    This is where we want to situate and emphasize the concept of \ef.
    Personal fabrication is becoming faster and easier, with a possible culmination being an effortless way to create matter and interact with it.
    If fabricating artifacts is an inherent, easy-to-access, and almost natural ability of humankind, what are the implications?
    One of them is the possible ephemerality of the resulting artifacts, which is an aspect that requires reflection from researchers and practitioners alike.
    
    The temporal aspect of artifacts, their existence, relocation, and lifetime were aspects of prior works.
    As described by Brand, aspects of buildings change with different rates (''shearing layers''): the site of a building is ''eternal'', services like wiring may last decades, the interior layout may change after few years, while ''stuff'' like furniture or other objects may move on a daily basis~\cite{brandHowBuildingsLearn1995}.
    Ephemeral artifacts may reside on a layer of change that is even more short-lived and temporary, than the arrangement of such appliances (''stuff'')~\cite{brandHowBuildingsLearn1995}, and come into existence for minutes while ceasing to exist as such a few moments later.
    D\"{o}ring et al. specified the term ''Ephemeral User Interfaces'' and the surrounding design space~\cite{doringDesignSpaceEphemeral2013}.
    Ephemeral interfaces are temporary, relying on deliberately chosen materials meant to last only a short time~\cite{kwonFugaciousFilmExploringAttentive2015,alakarppaBreathScreenDesignEvaluation2017}. 
    For \ef, in contrast, the \emph{interface} for it is lasting, while the resulting \emph{artifact} is not.
    That \emph{interface} can be any device or structure enabling access to material or potentially finished artifacts on demand.
    The \emph{artifact} is any shape made with the help of the interface to solve or address a -- potentially mundane -- task for a limited duration.
    Mark Weiser presented so-called ''pads'' as a counterpart to portable computers \cite{weiserComputer21St1991}.
    Pads were not meant to be permanent and personal, but rather ''scrap computers'' \cite{weiserComputer21St1991}.
    In the scenario of \ef, the \emph{interface} for it is, again, personal and is with the user all the time. 
    However, the artifacts users fabricate are likely to be treated as ''scrap artifacts'', fabricated in no time, and used for brief durations.
    Considering the \emph{material} to be dispensable and superfluous would be wasteful and unsustainable \cite{blevisSustainableInteractionDesign2007,vasquezEnvironmentalImpactPhysical2020}.
    Re-use, even in an enforced manner, is therefore crucial for \ef.
    
    \subsection{Ephemeral Problem-Solving and Bricolage}
        While \emph{manufacturing} ephemeral physical artifacts is currently rarely done by people, the act of \emph{ephemeral problem-solving} is far more prevalent.
        Human problem-solving can involve improvised solutions, which are supported by the availability of materials and tools around them and only last for a limited duration (i.e., as long as the requirement persists). 
        To mount a phone at a comfortable viewing angle, a stack of books may be (mis)used.
        To unscrew a slotted screw, a coin may replace the screwdriver, if it is not available.
        The resulting artifacts (coin-screwdriver, book-phone-mount) cease to be viewed as such as soon as their task has been fulfilled.
        None of these improvised solutions excels in functionality and efficiency, compared to specialized tools (a phone mount, a screwdriver), but may succeed in supporting users' tasks.
        These acts of ephemeral problem-solving can be considered to follow the notion of \emph{Bricolage}: acts of problem-solving with materials and tools that are at hand, which may not have been intended to be used for a particular challenge.
        Initially introduced by Claude Levi-Strauss, the concept of \emph{Bricolage}~\cite{levi-straussSavageMindPensee1966} gained relevance beyond the process of creating myths~\cite{johnsonBricoleurBricolageMetaphor2012,bakerImprovisingFirmsBricolage2003}. 
        Duymedjian und Rüling later specified the differences between ''\textit{bricoleur}'' and ''\textit{ingénieur}'' in organizations~\cite{duymedjianFoundationBricolageOrganization2010}. 
        For instance, the practice of the \textit{bricoleur} may operate ''with elements in stock''~\cite{duymedjianFoundationBricolageOrganization2010} and may not have clear outcomes~\cite{duymedjianFoundationBricolageOrganization2010}.
        The \textit{ingénieur}, however, embraces specifications and norms of processes~\cite{duymedjianFoundationBricolageOrganization2010}. 
        Ephemeral problem-solving primarily differs from acts of bricolage in the longevity of the solution: results of bricolage may, if desired, be kept for future (mis)use.
        These contrasting approaches have also been connected in HCI research~\cite{efratHybridBricolageBridging2016}.
        In this line of thought, it is possible to equate industry-grade fabrication to the \textit{ingénieur} and practices of craft to the \textit{bricoleur}. 
        Current trends in personal fabrication may be considered to be a bridging concept between \textit{bricoleur} and \textit{ingénieur}, adding precision without necessarily sacrificing improvisation, craft, and creativity.
        The ability to improvise, creating make-shift solutions and support tools is a core aspect of human nature. 
        However, since human beings lack an inherent fabrication ability (i.e., deploying matter to be used for basic mechanical tasks), this skill is still limited by the physical environment or a user's equipment being the provider of material and building blocks (i.e., the ''elements in stock''~\cite{duymedjianFoundationBricolageOrganization2010}) for the process of problem-solving.
        Being able to fabricate for, repair, and personalize the physical world ad-hoc, without a lengthy design process amplifies our ability to improvise and solve problems, introducing personal fabrication as a viable tool to more, even mundane, scenarios.
        A body-worn or even body-integrated fabrication device would support such a scenario further, by greatly reducing effort, access times, and complexity of the process, giving rise to a scenario such as \ef.

    \subsection{Fabrication as Human Augmentation}
        In nature, fabrication abilities are an integral part of the existence of some species~\cite{jacksonHowAnimalsFabricate2017,vonfrischAnimalArchitecture1974}.
        Apart from animals gathering material to fabricate nests (e.g., birds), there are species with \emph{integrated} fabrication abilities.
        They are outfitted with glands to produce materials used to fabricate structures and thereby solve specific mechanical tasks.
        Bees fabricate honeycombs to store resources and protect their offspring.
        Spiders are able to create webs out of different kinds of silk, to be able to catch prey or extend their sensing range.
        However, spiders use their silk oftentimes for much more than creating large (and often lasting) web structures. 
        Deinopidae create webs and attach them to their own front legs to be able to catch prey~\cite{vonfrischAnimalArchitecture1974}. 
        Hunting spiders use the silk as drag lines while hunting, create egg sacs, and even use it to pick up air currents and travel long distances \cite{pearceSpiderBallooningCrops2002}. 
        
        In popular culture, the character of Spider-Man depicts the transfer of such abilities to a human ''user'', applying these fabrication abilities to seemingly mundane tasks like affixing a camera to an arbitrary position or shoot strings of web to enable a fast form of locomotion.
        Notably, the created artifacts (e.g., the camera mount) merely last as long as they are needed, and can be considered a fictional act of \ef.
        This exemplifies how having inherent but easy and instant access to fabrication abilities, gives rise to fabrication scenarios that address highly specialized problems in-situ. 
        These scenarios are no obligatory occurrence, or unavoidable, but are acts of ephemeral problem-solving or bricolage, emerging from often small benefits.
        It furthermore demonstrates what we may consider a ''natural'' embedding of fabrication in a user's life.
        However, one may wonder who is cleaning up the consequences of these ''fabrication activities'', and to which degree the material can and is re-used, as the comic books often omit how this aftermath is dealt with\footnote{c.f. \emph{Damage Control}, which is an ''institution'' for comparable tasks.} or defer this to ''dissolvable'' artifacts.
    
    \subsection{Definitions and Shades of Ephemerality in Physical Artifacts}
        \label{subsec:eftable}
        \begin{table*}[ht!]
            \centering
            \includegraphics[width=\textwidth]{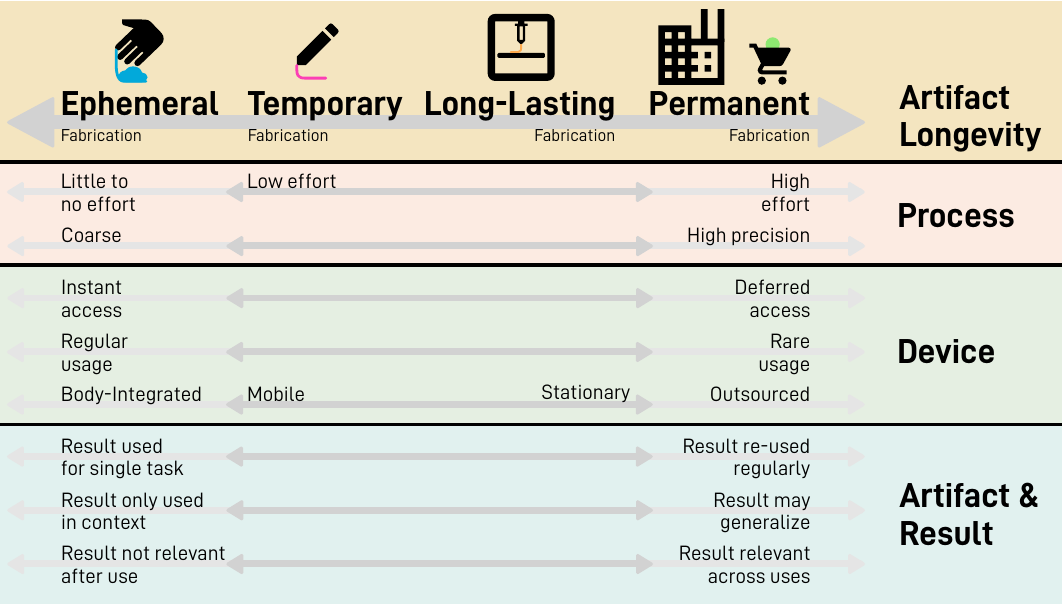}
            \caption{Dimensions of \ef in contrast to other fabrication scenarios. 
            While the artifact longevity is the core attribute separating \ef from other fabrication scenarios, other aspects can be derived from it:
            Effort, precision, access time, usage frequency, device integration, and the usage of the resulting artifacts. Specific systems may cover ranges of the longevity axis and deviate from sub-dimensions of a fabrication scenario (e.g., a \textbf{mobile} system for \ef).
            }~\label{tab:eph}
        \end{table*}
        Ephemeral artifacts are primarily objects with a short lifetime that require negligible effort to be created.
        We consider them to be a result of fabrication activities (analog or digital), in contrast to pure repurposing activities as mentioned earlier.
        However, the fabrication scenario (\emph{\ef}), which enables the creation thereof, has a set of prerequisites.
        These prerequisites differentiate \ef from other, more established, (personal) fabrication scenarios.
        Apart from being low- to zero-effort, the entire process should be as brief as possible, exhibiting a low temporal demand.
        Unlike industrial machinery or other CNC devices, ephemeral artifacts do not necessarily require precision in the fabrication process.
        Coarse shape-approximation, just enough to fulfill the \emph{functional} (e.g., a mechanical function) purpose of the artifact, is sufficient.
        If the environment or the task affords it, precision can be regained by applying methods like templating or molding, as described in section \ref{subsec:interactionpatterns}.
        Lastly, the means to fabricate should be accessible directly, immediately, and with as few indirections through disconnected tools (such as design software only available on stationary computers) as possible.
        
        The notion of \ef is surrounded by various dimensions where \system and prior work can be categorized and arranged.
        In the following, we present a set of dimensions we consider relevant for \ef (table \ref{tab:eph}).
        Ultimately, the extremes of these axes point to two distinct (theoretical) fabrication scenarios: pure permanent fabrication and pure ephemeral fabrication.
        While these terms primarily address the temporal dimension of the artifact (i.e., its service life), further aspects can be derived from them.
        The dimensions seen in table \ref{tab:eph} are interconnected.
        However, no rigid 1:1 mapping can be derived, as one may for instance use a stationary 3D-printer to fabricate \emph{ephemeral} artifacts, or use a 3D-printing pen to fabricate \emph{lasting} objects.
        Aspects like the required time or effort indicate suitability for a certain fabrication scenario (e.g., it may be unlikely, yet possible, for users to fabricate lasting objects with a 3D-printing pen).
        
        When viewed through a lens of \emph{current} personal fabrication devices and materials, a set of insights with respect to future \ef can be outlined.
        Within the scenario of \ef, certain dimensions emerge as relevant for the employed tools.
        Notably, the process should ideally be largely tool-less.
        Tools have to be carried, and therefore may not be available -- the core tool is the fabrication device itself and the immediate environment and its features (e.g., shapes to use as a mold).
        Likewise, the process has to be conducted ad-hoc (i.e., without lengthy preparation), be tangible, coarse, and as fast as possible while requiring a minimal amount of effort.
        The material used should also allow for changing aggregate states (i.e., malleable$\Leftrightarrow$solid) both for reuse, mechanical tasks, and interactive shaping.
        The process of (ephemeral) problem-solving is meant to yield a solution while the user is actively interacting with both the material and complementary workpieces (e.g., a phone and a table it is supposed to be mounted on).
        
        Prior work has covered some ranges of the dimensions relevant for \ef (table \ref{tab:eph}).
        However, the implicit goal was often to achieve a \emph{permanent} artifact, either with~\cite{shenRoCuModelIterativeTangible2013,nisserSequentialSupport3D2019,weichelReFormIntegratingPhysical2015,leenStrutModelingLowFidelityConstruction2017}, or without~\cite{yasuMOR4RHowCreate2016} explicit design iterations.
        This is in line with the outlook that personal fabrication empowers users with previously industry-only manufacturing capabilities, with the industry ideally producing long-lasting or permanent artifacts~\cite{baudischPersonalFabrication2017}.
        We refer to this direction as \pf, despite few consumer-grade artifacts reaching this degree of longevity. 
        \llf is what we consider to be a feasible outcome of contemporary, widely available, and accessible means for personal fabrication (e.g., 3D-printers). 
        Core difference is this particular availability of processes to enthusiast or hobbyist users.
        The artifacts may -- under ideal circumstances -- be permanent, but are often subject to prior iterations and may exhibit limitations due to material choice/availability.
        \tf aims for temporary artifacts, which may be created with lower precision than the aforementioned groups, subsequently requiring less time and effort.
        
        The vision of Mobile Fabrication by Roumen et al. may be considered an outstanding example, covering the range between \ef  and \tf, as it aims to be low-effort, the access times are ideally low and the resulting coarse artifact is meant to solve a single problem~\cite{roumenMobileFabrication2016}.
        However, the devices (pen or printer) are mobile and incur a delayed access time and the effort or process duration if tracing is employed is comparably high.
        Likewise, the usage is based on immediate \emph{need} (e.g., a problem), instead of a pure \emph{ability} to fabricate.
        The scenarios described in the work can be separated into \tf and \ef: a wrench made to repair a bike light is an ephemeral artifact and will likely not be reused a second time (\ef), while fabricated shirt buttons or shoelaces are not meant to cease to exist, but rather fulfill a function until a replacement (e.g., one generated through \pf) is found.
        Popfab by Peek and Moyer, is a desktop-grade device made portable, while retaining high effort and precision in the process~
        \cite{peekPopfabCasePortable2017}, covering \llf mainly, albeit in a portable context.
        The concept and system for ''Patching Physical Objects'' by Teibrich et al. implicitly covered the range from \tf, over \llf to \pf, making created artifacts alter-able, if changing requirements arise or the object breaks~\cite{teibrichPatchingPhysicalObjects2015}.
        Again, effort and time expenditure are high, while the device used is stationary and precise.
        The result is also explicitly aimed to last and be regularly used, therefore covering a range from \tf to \llf.
        WeaveMesh by Tao et al. emphasizes short iterations, with manual intervention by the designer~\cite{taoWeaveMeshLowFidelityLowCost2017}.
        While the single intermediary artifact is used only to refine the design, the final result is fabricated with high fidelity with the process similarly requiring relatively high effort (e.g., to assemble the intermediary artifact)~\cite{taoWeaveMeshLowFidelityLowCost2017}.
        BlowFab accelerates prototyping and allows for re-use of the artifacts, while additionally allowing the artifacts to deflate to save space~\cite{yamaokaBlowFabRapidPrototyping2017}.
        In the case of Sequential Support by Nisser et al., \emph{parts} of fabricated artifacts can be considered ephemeral or at least temporary~\cite{nisserSequentialSupport3D2019}.
        By allowing parts of the artifact to dissolve, partial replacement, additional temporary protection or time-controlled functions become possible~\cite{nisserSequentialSupport3D2019}.
        As before, the device is stationary, the process rightfully demands effort and may require larger amounts of time, while covering \tf to \llf. 
        It is crucial to consider the general notion of 'prototyping' artifacts during design processes.
        Prototypes are, inherently designed \emph{not} to be reused, but instead be artifact manifestations that drive a design process forward -- they are essentially \textit{ephemeral by design}. 
        This is appropriate for the time when merely hobbyists and researchers rely on personal fabrication, but does not scale to constant and widespread use.
        Few research approaches explicitly deal with this aspect of the process, and rather focus on the goal to be achieved, which is usually a fitting, well-executed and -fabricated object that \emph{lasts}.\\
    
        The following brief and mundane examples can be differentiated by the effort needed., especially in contrast to the ones presented in most aforementioned academic works.
        A user may want to mount a phone in a specific position to take a picture.
        If no tripod or a similarly specialized artifact is at hand, a different solution is required.
        The user may start improvising, engaging in \emph{Bricolage}, as is human nature: repurpose books, bags, sticks, and stones from her immediate vicinity to replace the function of ''holding a device in a steady angle and position''.
        To improve the handle of a hammer for better ''power grip''~\cite{napierPrehensileMovementsHuman1956,lewisDesignSizingErgonomic1993}, a user may start taping it with a non-slip material, if, and only if, it is at hand.
        If it is not, the entire process of ''adapting the grip'' is stretched to find or purchase the needed tape, which takes hours to days, instead of few moments.
        Alternatively, the user may add a more malleable material, grab the handle and thereby generate ridges that are unique to this type of grip and to the user's fingers.
        To reach a deeply recessed button (e.g., a reset button) a user may start searching for an arbitrary pointed and elongated object.
        Or, she could ''create'' it out of a malleable material which solidifies.
        All prior examples have in common that they are, to a degree solvable without the need for \ef. 
        In a sense, temporarily repurposing artifacts and materials from the users' physical context, can be considered an act of ephemeral problem-solving through \ef or craft. 
        The resulting configuration is equally temporary and ephemeral (i.e., it is supposed to last until the user changes the grip used to hold the hammer from a ''power grip'' to a ''precision grip''~\cite{napierPrehensileMovementsHuman1956}).
        However, if the environment is unable to provide the appropriate resources, fulfillment of the task through improvisation and repurposed objects becomes more complicated. 
        One can imagine various further application scenarios for ephemeral fabrication, some of which we depict in chapter \ref{sec:scenarios}.
        However, all of these scenarios follow a simple calculation: the creation of the artifact takes almost no time and effort and its benefits are worthwhile for the user. 
        The lower the effort, the more scenarios arise that are ''worth it'', even with minuscule benefits for the user, as the effort required to achieve the benefit is equally minuscule.
        This is what may emerge from ever-accelerating and -improving digital personal fabrication workflows, along with all associated positive and negative side effects.
        
        \subsection{Related Work}
            \label{subsec:rw}
            Having grounded our definitions of \ef through prior works, we also want to introduce a set of research efforts which depict a steady progress towards a future with \ef.
            The core dimensions we considered with the development of \ef were time and effort.
            Process steps like modeling or fabrication are equally valid for these considerations, in the sense that they are continuously reduced, changing our perception and application of the entire (personal) fabrication procedure.
            Both novice-friendly fabrication and portable fabrication were part of prior research and can be both considered attempts to address required effort for fabrication or attempts to broaden the contexts it may be employed in.
            The \ef scenario can be further embedded on the contexts of sustainability research (e.g.,~\cite{blevisSustainableInteractionDesign2007,barniUrbanManufacturingSustainable2019,wuUnfabricateDesigningSmart2020}), research driving the vision of widespread adoption of personal fabrication forward (e.g.,~\cite{baudischPersonalFabrication2017,gershenfeldDesigningRealityHow2017,stemasovRoadUbiquitousPersonal2021,stemasovEnablingUbiquitousPersonal2021}) and research in human augmentation through technology (e.g.,~\cite{shilkrotDigitalDigitsComprehensive2015,schmidtTechnologiesAmplifyMind2017,saraijiMetaArmsBodyRemapping2018}).

            \subsubsection{In-Situ, Mobile and Tangible Personal Fabrication}
                Research in personal fabrication focused heavily on the transfer and improvement of digital fabrication processes.
                Similarly to the previous group, projects focusing on in-situ or tangible personal fabrication are making it more accessible to a broader audience (e.g., people with no experience with CAD).
                Novice users have different requirements for the process of digital fabrication than industrial experts.
                To achieve better ease of use, tangible modeling was evaluated for the creation of spline-based geometry~\cite{shenRoCuModelIterativeTangible2013} or converted to a construction kit of pre-defined elements~\cite{leenStrutModelingLowFidelityConstruction2017}.
                These approaches increase ease-of use, while sacrificing a degree of achievable precision.
                To transfer traditional modeling to the digital domain in the enthusiast space, Jones et al. relied on Clay combined with functional elements~\cite{jonesWhatYouSculpt2016}.
                The resulting paradigm was called ''what you sculpt is what you get''~\cite{jonesWhatYouSculpt2016} and embraced analog modeling and the associated, tangible experience.
                Similarly, Weichel et al. aimed to combine traditional and digital fabrication with a novel, bidirectional workflow, which supported both kinds of in- and output~\cite{weichelReFormIntegratingPhysical2015}.
                One's own body provides complex but relevant geometric features for artifacts.
                This is relevant for hand-held or body-worn objects, which we consider a relevant task for \ef.
                Creating objects meant to interact with the body can for instance be done by measuring and modeling the body~\cite{gonzalezErgonomicCustomizedToolHandle2018},or fabricating right on its surface~\cite{gannonExoSkinOnBodyFabrication2016}.
                Attempts to transfer personal fabrication workflows towards mobile or nomadic use move the adoption of personal fabrication towards paradigms that give rise to more short-lived artifacts (c.f., figure \ref{fig:extrapolation}).
                These artifacts are coupled more closely to their physical context~\cite{peekPopfabCasePortable2017,roumenMobileFabrication2016,quitmeyerWearableStudioPractice2015,stemasovMixMatchOmitting2020}.
                For example, Quitmeyer and Perner-Wilson presented \emph{Wearable Studio Practice}, which focused on extraordinary environments for digital and analog craft used to develop new wearable technologies~\cite{quitmeyerWearableStudioPractice2015}.
        
                \ef can be considered a culmination of the aforementioned research trends: personal fabrication that requires little effort, is highly mobile, but still yields highly personalized artifacts that are created within their physical context (both for a limited timespan and a unique requirement).
                The aforementioned works influenced the development of \ef and \system in two ways: 1) they depict a steady progress towards a wider userbase for personal fabrication through simplification and acceleration and 2) they explore novel modalities beyond screen-based CAD for digital fabrication.
                
            \subsubsection{Sustainability in Fabrication and HCI}
                Sustainability considerations have also played a role in personal and digital fabrication research.
                Vasquez et al. surveyed a set of papers involving physical prototyping, discussing both energy consumption of machines and material impact~\cite{vasquezEnvironmentalImpactPhysical2020}.
                While prototyping is highly valuable, it is currently not as widely employed yet.
                If fabrication permeates more contexts and user groups, a reconsideration of this is likely needed.
                In ''Making 'Making' Critical'', Cindy Kohtala discussed how sustainability considerations are connected to social and ideological considerations in the context of Fab Labs as organized entities enabling personal fabrication~\cite{kohtalaMakingMakingCritical2017}.
                Dew and Rosner likewise focused on makerspaces, exploring how the treatment of material as waste can be changed and made more thoughtful and deliberate~\cite{dewDesigningWasteSituated2019}.
                % Fab
                Wu and Devendorf explored how smart textiles can be designed with disassembly in mind~\cite{wuUnfabricateDesigningSmart2020}, without deferring this issue to a point in time where fabrication has already happened.
                Lazaro and Vega presented Myco-Acessories, combining reusable electronic components with compostable mycelium for wearable artifacts~\cite{vasquezMycoaccessoriesSustainableWearables2019,vasquezPlasticBiomaterialsPrototyping2019}.
                In contrast, Song and Paulos explored the act of ''Unmaking'', which discusses change and interaction after an object was seemingly ''finished''~\cite{songUnmakingEnablingCelebrating2021}.
                Choi and Ishii presented Therms-Up! as a way of creating inflatables from plastic bags on commodity 3D-printing hardware ~\cite{choiThermsUpDIYInflatables2021}, which is an upcycling process.
                ReFabricator by Yamada et al. is a tool with integrates arbitrary objects in a digital fabrication process, instead of relying on a 3D-printer for the entire artifact~\cite{yamadaReFabricatorIntegratingEveryday2016}.
                Scrappy by Wall et al. takes a similar direction, allowing users to replace infill of 3D-printed objects with prior iterations, discarded tools, or other waste~\cite{wallScrappyUsingScrap2021}.
                This reduces material use and accelerates the printing process at the same time.
                
                The aforementioned works are outstanding examples of how sustainability is both explored an embedded within user-facing personal fabrication systems and contexts.
                We firmly believe that such systems are crucial for a positive manifestation of \ef in the future.
                With \system, we embrace a certain circularity of material, over a net positive use of it.
                We further want to elicit new considerations about a future, where personal fabrication has indeed become as ubiquitous as computing itself~\cite{stemasovRoadUbiquitousPersonal2021}.

            \subsubsection{Human Augmentation}
                Symbiosis between human and machine was topic in HCI research for decades, even prior to the advent of wearable computing~\cite{lickliderManComputerSymbiosis1960}.
                The directions can be coarsely separated into cognitive augmentation~\cite{rhodesRemembranceAgentContinuously1996} and mechanical augmentation~\cite{leighMorphologyExtensionKit2018} of human abilities.
                The latter was the initial context of our work.
                A way to augment human abilities is the conceptual transfer of capabilities from the animal world.
                This includes sensory modalities, like being able to perceive infrared/heat signatures \cite{abdelrahmanSeeFireEvaluating2017} or sense objects in the environment with the help of ultrasound distance sensors~\cite{mateevitsiSensingEnvironmentSpiderSense2013}.
                Likewise, changes to body morphology transferred from other species, like the addition of a tail~\cite{nabeshimaProstheticTailArtificial2019} can be considered nature-inspired, biomimetic augmentations.
                Other (functional) alterations of body schema include supernumerary fingers~\cite{huHandDevelopmentKit2017} or other limbs~\cite{al-sadaOrochiInvestigatingRequirements2019,saraijiMetaArmsBodyRemapping2018,sasakiMetaLimbsMultipleArms2017}.
                The integration of tools into or onto the human body in general covers various concepts.
                This includes finger-worn devices, as presented by Shilkrot et al. with FingerReader.
                It is an assistive device meant to help with the perception of text for people with visual impairments~\cite{shilkrotFingerReaderWearableDevice2015}.
        
                We would frame always-accessible personal fabrication as a path and component of human augmentation, inspired by nature. 
                This amplifies their ability for (ephemeral) problem-solving or bricolage by ideally immediately providing always-available material, feasible for addressing mechanical tasks and challenges.
                It likewise lowers the threshold to apply personal fabrication to mundane, short-lived challenges faced by users -- which, in turn, gives rise to scenarios such as \ef.
                Whether an augmentation like \system would be perceived as a tool or as an ''inherent ability'' depends on how close the device would be coupled to the user.

\section{Prototype}
\label{sec:prototype}
    \system is a proof-of-concept implementation depicting the speculative vision of \ef.
    The diegetic prototype was presented through the narrative in section \ref{fig:narrative}.
    The following paragraphs describe the development and details of the physical prototype, as implementable with current means and materials.
    We consider our implementation of \system a very early version of a device for \ef, albeit one that already fulfills some aspects that enable \ef.
    As a proof-of-concept implementation, it provides the most fundamental functionality: extrusion/addition of a malleable material.
    The development and usage of \system enabled us to explore the concept of \ef.
    We do not consider the system to be the core contribution of this work, but rather an indispensable part of the \emph{process} (c.f., section \ref{sec:designprocess}), which depicts and supports our thought and design procedure.
    In its current state, \system is a brief step back from the \textit{digital} in personal fabrication.
    A core benefit -- digital \emph{precision} --  becomes less relevant to enable fast design and fabrication processes.
    Digital components are nevertheless used to couple fabrication device and human to enable material access and reduce time and effort. 
    This digital precision may re-emerge later through more sophisticated wearable systems, while compensating for the skill floor tangible modeling with PCL exhibits.
    Implementing \system through the means available to us right now, made the speculative narrative of \ef a tangible, explorable experience. 

        \begin{figure}[ht!]
            \centering
            \includegraphics[width=\minof{\columnwidth}{0.75\textwidth}]{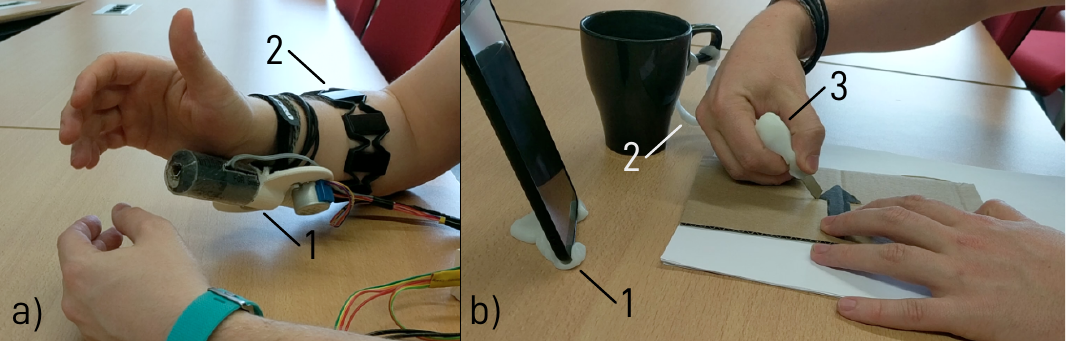}
            \caption{a) A user wearing the \system prototype. The extruder (1) and the EMG sensor (2) are visible. b) Results made with the prototype in use: a phone mount (1), a stabilizing extension to a mug handle (2), and an ergonomic blade grip used to cut cardboard (3).}
            \label{fig:systemAndResults}
        \end{figure}

    \subsection{Implementation}
    \label{subsec:prototype}
        For our implementation, we settled on a \emph{wearable} system to extrude malleable PCL (polycaprolactone), while coupling the device's function to the user's body signals through electromyography (EMG).
        We consider both the material choice and the choice to build Draupnir as a wearable, not a body-integrated system, to be a feasible compromise in light of current, practical means.
        Future systems, as outlined in section \ref{subsec:narrative}, may decrease access times and increase precision further.
        Prior research has used HDPE (high-density Polyethylene)~\cite{oxmanFreeform3DPrinting2013}, wax \cite{pengDCoilHandsonApproach2015}, ''jumping clay'' (water-based) \cite{gannonExoSkinOnBodyFabrication2016} and ''TecClay'' \cite{weichelReFormIntegratingPhysical2015}, apart from ABS or PLA.
        After initial explorations, we settled on PCL.
        It was also employed in recent works enabling post-print shape-change~\cite{koDesigningMetamaterialCells2021}.
        Core benefits of the material are skin-safety, sufficient time to model until the material solidifies, rigidity when thick and flexibility when thin.
        PCL can also be re-heated to be altered and, most importantly, reused for new artifacts. approach with a relatively wide nozzle.
        
\begin{figure}[t]
            \includegraphics[width=\minof{\columnwidth}{0.75\textwidth}]{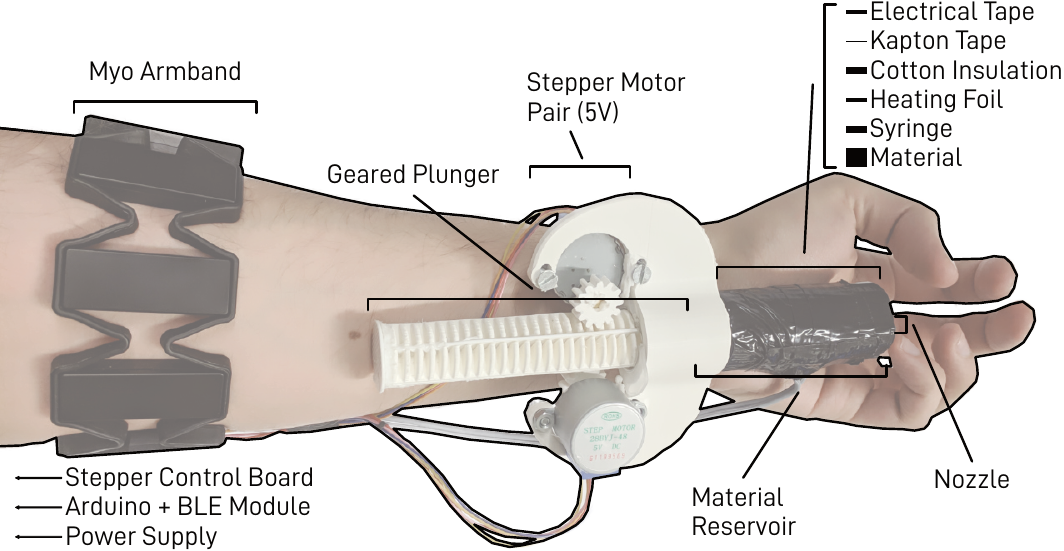}
            \caption{Core components of the \system prototype: a syringe-based paste extruder, driven by two stepper motors and controlled via an EMG-sensing arm-band. To protect the wearer from heat and to reduce the power draw needed to sustain the material's malleable state, multiple layers of insulation are present.}~\label{fig:setup}
        \end{figure}

        The resulting prototype is a \emph{wearable} paste extruder~\cite{amzaPasteExtruderHardware2017,puschLargeVolumeSyringe2018} for Polycaprolactone, which uses a syringe with a motorized piston to push out the preheated, soft material (see figure \ref{fig:systemAndResults}). 
        Movement of the nozzle if offloaded to the wearer, while heating and extrusion remain digitally-controlled.
        Concepts like ''Being the Machine'' by Devendorf and Ryokai took this approach even further~\cite{devendorfBeingMachineReconfiguring2015}.
        A 12V adhesive polyester heating foil (12W) covers the tube's front part (7cm).
        The model used is rated to reach 80\textdegree C under ideal circumstances.
        In our configuration, the (uncovered) foil reached approximately 65\textdegree C maximum.
        To ensure that no heated area comes in contact with the users' skin, we covered the surface with cotton insulation.
        The tip was insulated in a similar fashion.
        The resulting surface stayed below 40\textdegree C, even after prolonged use.
        As the heating pad only covers the sides of the syringe, the tip is heated less well.
        This was compensated by a thin sheet of thermally conductive copper around it.
        This technical setup can be seen in figure \ref{fig:setup}.
        
        The syringe is a single-use plastic model and is rated to contain 60ml of liquid, while withstanding the heat it is exposed to.
        \system can provide approximately 20g (or 28.5cm$^3$) of material, extruding it through a nozzle of 1 cm diameter.
        This ensures quicker extrusion and an amount of matter that can be conveniently shaped in a brief moment.
        When heated, the material may be retracted back into the syringe.
        The piston/plunger was lubricated with synthetic PTFE grease to reduce friction between the parts.
        Two stepper motors (28BYJ-48, 5V) are affixed at the sides of the rack, stabilizing the plunger from 2 sides and driving it.
        Similarly to \cite{leighBodyIntegratedProgrammable2016}, we used a Myo\footnote{\url{support.getmyo.com/hc/en-us}, Accessed: 25.6.2021} gesture control armband to receive the user's input.
        This tightly couples the device with the user's body signals, as an attempt to move farther away from the perception of ''a tool'' towards ''an \emph{ability}''.
        If the users move their wrist outward, the extrusion is enabled.
        This couples fabrication with the users' physiology more closely than previous approaches, while reducing access time and, ideally, effort.
        To connect it directly to the Arduino, the MyoBridge\footnote{\url{github.com/vroland/MyoBridge}, Accessed: 25.6.2021} Library/Firmware was used.
        Building \system as a working artifact enabled us to explore and develop the scenarios and interaction patterns presented in the following sections.
        
    \subsection{Usage Scenarios of \ef}
        \label{sec:scenarios}
    In the following, we present usage scenarios that we were able to generate and explore through the lens of \system, as implemented with current means. 
    These are certainly not \emph{all} the potential scenarios of \ef, but rather the ones we were able to explore and verify. 
    While they are a subset, they demonstrate a certain usefulness of \ef, despite an absence of high precision or industry-grade visual appeal.
    This usefulness is demonstrated through coarse artifacts, which can "solve" small challenges for -- usually -- brief timespans.
    The artifacts are quick to create and are seemingly mundane tasks where the effort currently needed to solve them (e.g., by employing digital design and fabrication processes) outweighs the necessity to address them.
    They furthermore assume that the ephemeral artifact created is likely to be created for the duration of its usage and perish afterward, leaving the raw material for other, equally mundane but functional, creations.
    Other scenarios could be generated by using alternative implementations of \system (e.g., carrying a material like clay or Sugru in one's pocket, or a hot glue gun in one's bag), which would result in alternative usage scenarios (based on aspects like access time or material properties) but may still fall under the umbrella of \ef. 
    At the same time, all scenarios inherently suffer from the same potential issues (i.e., thoughtless creation of single-use trash) of \ef in the context of sustainability and reuse.
    This is particularly the case if these scenarios are imagined as activities executed by a large userbase in numerous contexts.\\
    
    \begin{figure*}
        \centering
        \includegraphics[width=\textwidth]{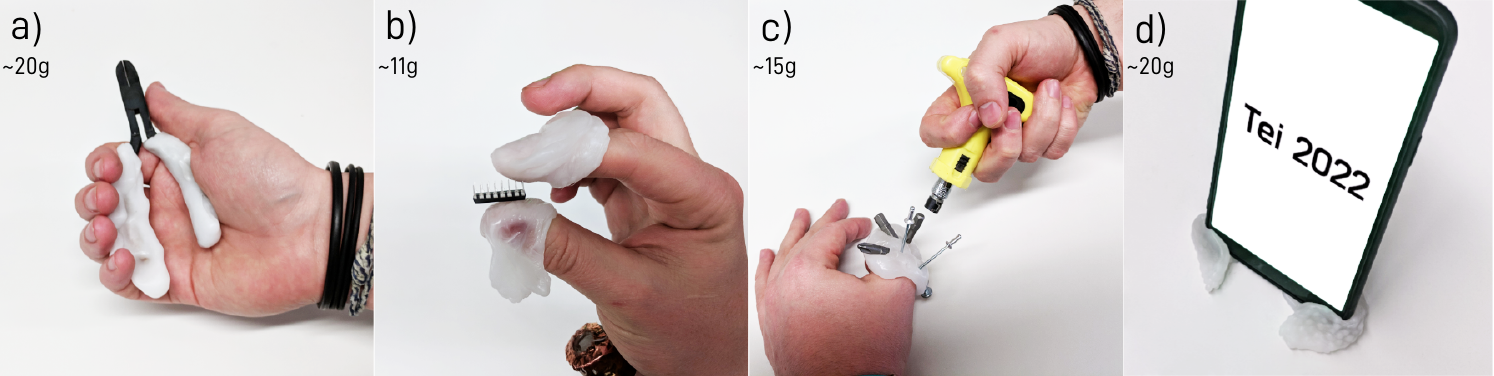}
        \caption{Exemplary set of scenarios for \ef with \system and the amount of material used (out of 20g it provides): a) Personalized ridges are added to a bare metal handle to improve grip, b) A protective film is added to one's fingers to be able to grasp sharp objects, c) An additional, mount for tools and parts is added to the hand, allowing for quick access to the carried objects, d) A temporary mount for a phone is made, to prop it up at a comfortable viewing angle, or to serve as a tripod.}~\label{fig:scenarios}
    \end{figure*}

    The process of creating an ephemeral artifact may be related to existing artifacts in the environment \cite{roumenMobileFabrication2016,chenEncore3DPrinted2015}, but also the users' bodies~\cite{gannonTactumSkinCentricApproach2015,gannonExoSkinOnBodyFabrication2016}.
    The use of templates taken from the users' physical context is likely to be an inherent part of the design and modeling process for any device enabling \ef, as it not only compensates for the absence of precision, but also reduces the effort to ''model'' or replicate features of objects or the users' bodies.
    The artifacts created cover a gradient of body-proximity, ranging from remote artifacts (''off body'', such as mounts for objects), over artifacts that interact with the body (''for body'', such as coverings of objects to grasp) and artifacts that are body-worn (''on body'', such as rings).\\ 
    
    \subsubsection{On-Body \ef}
    When retrofitted with an always available gland for fabrication, users can add a protective film to body regions (e.g., a thimble).
    The material used by \system (PCL) is rigid enough to protect from sharp edges.
    When required, users may create protective layers of varying thickness, for instance, to grasp and manipulate sharp objects with their fingers (figure \ref{fig:scenarios}b) or add mount points for other objects (\ref{fig:scenarios}c).
    Potentially, tailored, temporary casts can also be developed ad-hoc (i.e., right after the moment of injury) and fit to the user's proportions, remaining relevant until a better and more permanent cast is made (i.e., by a professional), which ventures in the realm of \tf.\\
    
    \subsubsection{For-Body \ef}
    A body-worn fabrication device for \ef also allows for a more refined, temporary customization experience.
    With \system and other future \ef devices, each physical artifact can be altered to a certain degree, as users are handed a personal tool for shape-change.
    An example is \ef to improve ergonomics temporarily.
    The ergonomic design of handles, for instance, is deliberately made for the general public.
    Diameters of handles and their orientation are specifically made to cover a broad range of hand sizes and proportions~\cite{lewisDesignSizingErgonomic1993}.
    It is discouraged to add ridges for fingers, as they are likely to make handling worse for the majority of users~\cite{lewisDesignSizingErgonomic1993}.
    For instance, it is possible to temporarily alter tool handles in place to provide better ergonomics and usage comfort.
    An example is shown in figure \ref{fig:scenarios}a, where a bare metal handle is covered and improved with personalized ridges for the duration of use.\\
    
    \subsubsection{Off-Body \ef}
    Apart from closely body-related \ef activities, a wearable fabrication system for \ef also supports other alterations to the physical world. 
    These are more closely related to established digital fabrication processes, but benefit from a quick process of tangible modeling.
    Creating mounts and fixtures for various objects is a general use-case made easier by easily accessible, low-effort \ef devices.
    Usually, when relying on CAD, the referential use of an object involves either scanning, measuring, or recreating the artifact in a different fashion~\cite{mahapatraBarriersEndUserDesigners2019}.
    This is not necessary when the object and the malleable material are allowed to interact. 
    For example, it is possible to create a phone mount, to affix the phone at a comfortable viewing angle (figure \ref{fig:scenarios}d).
    The mount itself can be directly manipulated, with the effects being directly perceptible.\\

    \label{subsec:interactionpatterns}
    Actively using \system allowed us to uncover a set of more specific interaction patterns relevant for \ef.
    We consider the following 4 patterns to be crucial for \ef, as they apply to most, if not all, tasks where the future artifact is meant to be ephemeral and made effortlessly:\newline
    \textbullet~\emph{\textbf{The Body as a Template}}: 
    any user of a body-worn fabrication system carries an essential reference with them -- their own body.
    This improves any tailoring process and allows to create custom-fit artifacts.
    These may include attachable artifacts, alterations to existing objects, but also modifications to body morphology.\newline
    \textbullet~\emph{\textbf{The World as a Template}}: 
    the physical environment can provide the most necessary measures and molds for the artifacts to be created. 
    Precise measurements can be omitted and replaced with tangible manipulation of the material and the complementary workpiece it has to interact with.\newline
    \textbullet~\emph{\textbf{Combinatorial/Augmented Fabrication}}: 
    Existing artifacts can easily be made part of a newly fabricated one, by including them as templates or integrating them in the new artifact~\cite{ashbrookAugmentedFabricationCombining2016,yungPrinty3DInsituTangible2018,chenEncore3DPrinted2015}.
    Molding combines the artifact with replicated features of another one -- parts of existing objects may be included in a new design (e.g., a screwdriver bit or a saw).
    Likewise, attached artifacts are also a result of such a combinatorial or augmented~\cite{ashbrookAugmentedFabricationCombining2016} fabrication.\newline
    \textbullet~\emph{\textbf{Ephemeral Fabrication \underline{dictates} Re-Use}}: 
    the speculative future of \ef can be considered an unsustainable one.
    If personal fabrication is effortless and quick, permeating through contexts it is currently not applicable to, it either leads to copious amounts of waste; or it leads to more thoughtful, re-use-focused processes that do not treat the consequences of creating physical artifacts as an afterthought.
    The latter is what we aimed to demonstrate with \system and its constraint of limited available material, ensuring and, most importantly, enforcing re-use, instead of deferring it to other entities and later points in time.
    
    The first 3 patterns are focused on (but not limited to) \system: they allow users to compensate for the disadvantages of a system that does not generate precise shapes.
    Furthermore, they try to make \ef~''work'' as a feasible fabrication method (for short-lived everyday tasks).
    It is crucial to consider that they demonstrate a certain feasibility of such a device, \emph{regardless of how re-use is embedded or considered}.
    This last pattern, re-use, does not only apply to systems aimed at \ef, but rather all personal fabrication systems.
    In particular, it becomes highly relevant when personal fabrication systems are adopted by more people in more contexts, while still generally deferring sustainability considerations to material scientists, or deferring them to later points in time (e.g., biodegradation of PLA).

\section{The Role of Sustainability in Ubiquitous Personal Fabrication}
\label{sec:limitations}
    With \system, we, in part, paint an optimistic outlook on \ef: one that embraces ephemerality, while (ideally) enforcing re-use.
    The easier, faster, and effortless personal fabrication becomes, the more adoption it will see.
    Adoption does not only apply to users and user groups, but also the contexts and tasks it is employed and considered relevant for.
    Paradigms like wearable fabrication or body-integrated fabrication (analogous to wearable computing) are likely to drive this adoption further, up until a point where personal fabrication is a ubiquitous concept~\cite{stemasovRoadUbiquitousPersonal2021}, which entails both opportunities (e.g., augmented problem-solving capabilities) and dangers (e.g., a throw-away society with even more means to create single-use artifacts).
    This ubiquitous adoption of personal fabrication, especially through methods that enable users to reach their goals quickly, calls for a consideration of sustainability~\cite{baudischPersonalFabrication2017}: one that embeds it in interactions a priori, instead of considering it after a tool was developed or a fabrication activity has already happened.
    Regardless of the means (devices, materials, software) enabling ubiquitous personal fabrication, \ef remains a possible future fabrication scenario.
    \ef is unlikely to be the only expression of personal fabrication in the future.
    It can be considered a complementary method to more established means for ''attaining physical artifacts'' (e.g., personal fabrication, shopping).
    However, \ef may become a \textbf{dominant} paradigm for personal fabrication, if it culminates in outstandingly fast and effortless design and fabrication methods.
    This is especially the case if a low effort is combined with numerical precision, which was not the case for our implementation of \system.
    For this particular case of personal fabrication becoming a ubiquitous phenomenon, re-use, embracing artifact ephemerality, and low-effort processes are crucial to consider, enable, and potentially enforce.
    \ef may take a positive or a negative expression. 
    This depends on \emph{how} sustainability is embedded in systems and processes (by design or as a consideration afterward).
    The \textit{\ef Dystopia} consists of widespread (users and contexts) personal fabrication, with no re-use and no constraints imposed on the process.
    The \textit{\ef Utopia} likewise consists of widespread personal fabrication, where re-use of ephemeral and lasting artifacts is considered and embedded early on while being \emph{just as easy and effortless as their fabrication}.
    The future personal fabrication is moving towards may not be focused on re-use, with design, iteration, and fabrication for \emph{lasting} artifacts taking center stage of current HCI-focused research.
    
    \subsection{Negative Embedding of Ephemerality}
        If we are dealing with negligible effort and time consumption, it is possible that artifacts fulfilling their function lose any value (and inhibition to engage in personal fabrication) they receive from the process itself.
        This is a dystopian fiction, where any person on earth has access to fabrication devices, which easily enable fabrication, but not reuse (or consider reuse to be ''optional'').
        Few modeling tools aim for reuse by design, as their goal is to create lasting artifacts.
        This is connected to one of the goals of personal fabrication: to make industry-grade manufacturing available to any potential user~\cite{gershenfeldDesigningRealityHow2017,baudischPersonalFabrication2017}.
        However, reconsidering this goal and the means that enable it is required to understand and evaluate its consequences.
        A future of fleeting, ephemeral physical artifacts that are created on-demand, and only recycled if a user desires so, is incompatible with one where large user groups have access to the tools for doing so.

    \subsection{Positive Embedding of Ephemerality}
        We consider this work to be a design probe into the space of future personal fabrication and its impact on sustainability and the environment.
        \system and the developed interaction patterns mostly belong into a realm of utopian personal fabrication, which relies on reuse as an inherent component, while enabling and encouraging the user to employ personal fabrication in arbitrary, often mundane use cases.
        Ubiquitous personal fabrication is a conceivable future and requires that the issues of it, evident in current trends, are discussed and addressed in current research, \emph{before} it becomes a ubiquitous phenomenon~\cite{stemasovRoadUbiquitousPersonal2021}.
        While we cannot precisely anticipate the impact of widespread adoption of personal fabrication, we can explore potential futures through diegetic and physical prototypes as an attempt to resolve the Collingridge Dilemma we face~\cite{collingridgeSocialControlTechnology1982}.
        Designing fabrication techniques to incorporate disassembly as demonstrated by Wu and Devendorf~\cite{wuUnfabricateDesigningSmart2020} is one way to \emph{positively} address artifact ephemerality while still embracing iterations and prototypes.
        Similarly, Song and Paulos embrace the act of ''unmaking'', focusing on processes starting from a seemingly ideal, finished artifact instead of initiating from (raw) material~\cite{songUnmakingEnablingCelebrating2021}.
        A way to nudge this consideration further into the view of researchers and practitioners are impact statements which were suggested before~\cite{shneidermanHumanValuesFuture1999,palCHI4GoodGood4CHI2017} and can be considered for any type of societal and environmental impact\footnote{As introduced for NeurIPS in 2020: \url{https://neuripsconf.medium.com/getting-started-with-neurips-2020-e350f9b39c28}, Accessed: 5.7.2021}.
        The limited amount and type of material we chose as a core for the \system prototype already provides the necessary constraints for embedded re-use: the produced artifact \emph{can} and \emph{should} be re-shaped and re-used (or retracted/fed back into the reservoir).
        However, it is similarly crucial to consider the energy impact, in addition to the material impact of a fabrication system~\cite{vasquezEnvironmentalImpactPhysical2020}, which was less the case for \system.
        Despite limitations of the prototype (mobility, cumbersome retraction mechanism), the actual implementation of \system allowed a more practical exploration of \ef and associated scenarios.
        It furthermore allowed for an exploration of techniques and materials available to us at the moment, which can be reused for further prototyping of comparable systems.
        \system was our vehicle to explore, understand, and critically reflect on the \ef scenario.\\

\section{Conclusion}
\label{sec:conclusion}
    \ef is by far not the first or only work that argues for sustainability in personal fabrication~\cite{wuUnfabricateDesigningSmart2020,vasquezEnvironmentalImpactPhysical2020,oxmanFreeform3DPrinting2013,teibrichPatchingPhysicalObjects2015,dewDesigningWasteSituated2019}. 
    However, we present an additional argument to not only consider sustainability as an afterthought but to put it at the core of future, user-facing personal fabrication tools by deeply embedding it in the interactions with fabrication devices or systems.
    We also present a way how constant, pervasive fabrication of mundane artifacts does not have to be an exclusively negative outlook.
    It does not have to entail the constant creation of trash but rather be an augmentation of human problem-solving capabilities.
    
    We presented the \ef scenario -- a speculative future where the sheer ability and effortlessness to quickly fabricate coarse yet functional artifacts in any given context empowers users to solve mundane tasks and inconveniences with the means of personal fabrication.
    We specified \ef as a fabrication scenario that may emerge from continuous improvements in the processes required for personal fabrication, reducing their complexity, stationarity, time, and effort needed to fulfill a given unique personal desire or requirement.
    \ef can be set apart from other interaction scenarios, where the goal is usually to create \emph{lasting artifacts} or address \emph{urgent demands}, which currently necessitates effort and time.
    \ef does not necessarily have to be a dystopian outlook.
    A \textit{positive} future expression of \ef entails that personal fabrication became a ubiquitous technology~\cite{stemasovRoadUbiquitousPersonal2021,gershenfeldDesigningRealityHow2017}, while being adapted to the new requirements and responsibilities of material usage.
    This speculative future focuses on the sustainability of personal fabrication, \emph{integrating} re-use in a closed-loop of material~\cite{vogelCircularHCITools2020,zhuRealizationCircularEconomy2021}.
    A \textit{negative} future expression of \ef, on the other hand, emerges from merely faster and more accessible tools for personal fabrication without the considerations of their impact (if genuinely adopted by a majority of the population).
    Whether and how exactly this future may emerge is in the hands of researchers and practitioners currently working on novel, ever-improving systems, devices, and processes.
    We strongly believe that re-use and associated concepts should be woven into novel personal fabrication workflows, materials, and systems.
    With fabrication processes becoming increasingly portable, fast, and effortless, sustainable use of artifacts and material becomes indispensable \emph{before} the means for it become pervasive technologies.

\begin{acks}
    We thank all our anonymous reviewers who have helped shape and improve the paper.
\end{acks}

\bibliographystyle{ACM-Reference-Format}
\bibliography{main}


\begin{thebibliography}{97}

%%% the Web address.

\ifx \showCODEN    \undefined \def \showCODEN     #1{\unskip}     \fi
\ifx \showDOI      \undefined \def \showDOI       #1{#1}\fi
\ifx \showISBNx    \undefined \def \showISBNx     #1{\unskip}     \fi
\ifx \showISBNxiii \undefined \def \showISBNxiii  #1{\unskip}     \fi
\ifx \showISSN     \undefined \def \showISSN      #1{\unskip}     \fi
\ifx \showLCCN     \undefined \def \showLCCN      #1{\unskip}     \fi
\ifx \shownote     \undefined \def \shownote      #1{#1}          \fi
\ifx \showarticletitle \undefined \def \showarticletitle #1{#1}   \fi
\ifx \showURL      \undefined \def \showURL       {\relax}        \fi
% invisible to TeX
\providecommand\bibfield[2]{#2}
\providecommand\bibinfo[2]{#2}
\providecommand\natexlab[1]{#1}
\providecommand\showeprint[2][]{arXiv:#2}

\bibitem[\protect\citeauthoryear{Abdelrahman, Knierim, Wozniak, Henze, and
  Schmidt}{Abdelrahman et~al\mbox{.}}{2017}]%
        {abdelrahmanSeeFireEvaluating2017}
\bibfield{author}{\bibinfo{person}{Yomna Abdelrahman}, \bibinfo{person}{Pascal
  Knierim}, \bibinfo{person}{Pawel~W. Wozniak}, \bibinfo{person}{Niels Henze},
  {and} \bibinfo{person}{Albrecht Schmidt}.} \bibinfo{year}{2017}\natexlab{}.
\newblock \showarticletitle{See {{Through}} the {{Fire}}: Evaluating the
  {{Augmentation}} of {{Visual Perception}} of {{Firefighters Using Depth}} and
  {{Thermal Cameras}}}. In \bibinfo{booktitle}{\emph{Proceedings of the 2017
  {{ACM International Joint Conference}} on {{Pervasive}} and {{Ubiquitous
  Computing}} and {{Proceedings}} of the 2017 {{ACM International Symposium}}
  on {{Wearable Computers}}}} \emph{(\bibinfo{series}{{{UbiComp}} '17})}.
  \bibinfo{publisher}{{ACM}}, \bibinfo{address}{{New York, NY, USA}},
  \bibinfo{pages}{693--696}.
\newblock
\showISBNx{978-1-4503-5190-4}
\urldef\tempurl%
\url{https://doi.org/10.1145/3123024.3129269}
\showDOI{\tempurl}


\bibitem[\protect\citeauthoryear{{Al-Sada}, H{\"o}glund, Khamis, Urbani, and
  Nakajima}{{Al-Sada} et~al\mbox{.}}{2019}]%
        {al-sadaOrochiInvestigatingRequirements2019}
\bibfield{author}{\bibinfo{person}{Mohammed {Al-Sada}}, \bibinfo{person}{Thomas
  H{\"o}glund}, \bibinfo{person}{Mohamed Khamis}, \bibinfo{person}{Jaryd
  Urbani}, {and} \bibinfo{person}{Tatsuo Nakajima}.}
  \bibinfo{year}{2019}\natexlab{}.
\newblock \showarticletitle{Orochi: Investigating {{Requirements}} and
  {{Expectations}} for {{Multipurpose Daily Used Supernumerary Robotic
  Limbs}}}. In \bibinfo{booktitle}{\emph{Proceedings of the 10th {{Augmented
  Human International Conference}} 2019 on - {{AH2019}}}}.
  \bibinfo{publisher}{{ACM Press}}, \bibinfo{address}{{Reims, France}},
  \bibinfo{pages}{1--9}.
\newblock
\showISBNx{978-1-4503-6547-5}
\urldef\tempurl%
\url{https://doi.org/10.1145/3311823.3311850}
\showDOI{\tempurl}


\bibitem[\protect\citeauthoryear{Alak{\"a}rpp{\"a}, Jaakkola, Colley, and
  H{\"a}kkil{\"a}}{Alak{\"a}rpp{\"a} et~al\mbox{.}}{2017}]%
        {alakarppaBreathScreenDesignEvaluation2017}
\bibfield{author}{\bibinfo{person}{Ismo Alak{\"a}rpp{\"a}},
  \bibinfo{person}{Elisa Jaakkola}, \bibinfo{person}{Ashley Colley}, {and}
  \bibinfo{person}{Jonna H{\"a}kkil{\"a}}.} \bibinfo{year}{2017}\natexlab{}.
\newblock \showarticletitle{{{BreathScreen}}: Design and {{Evaluation}} of an
  {{Ephemeral UI}}}. In \bibinfo{booktitle}{\emph{Proceedings of the 2017 {{CHI
  Conference}} on {{Human Factors}} in {{Computing Systems}}}}
  \emph{(\bibinfo{series}{{{CHI}} '17})}. \bibinfo{publisher}{{ACM}},
  \bibinfo{address}{{New York, NY, USA}}, \bibinfo{pages}{4424--4429}.
\newblock
\showISBNx{978-1-4503-4655-9}
\urldef\tempurl%
\url{https://doi.org/10.1145/3025453.3025973}
\showDOI{\tempurl}


\bibitem[\protect\citeauthoryear{Amza, Zapciu, and Popescu}{Amza
  et~al\mbox{.}}{2017}]%
        {amzaPasteExtruderHardware2017}
\bibfield{author}{\bibinfo{person}{Catalin~Gheorghe Amza},
  \bibinfo{person}{Aurelian Zapciu}, {and} \bibinfo{person}{Diana Popescu}.}
  \bibinfo{year}{2017}\natexlab{}.
\newblock \showarticletitle{Paste {{Extruder}}\textemdash{{Hardware
  Add}}-{{On}} for {{Desktop 3D Printers}}}.
\newblock \bibinfo{journal}{\emph{Technologies}} \bibinfo{volume}{5},
  \bibinfo{number}{3} (\bibinfo{date}{Sept.} \bibinfo{year}{2017}),
  \bibinfo{pages}{50}.
\newblock
\urldef\tempurl%
\url{https://doi.org/10.3390/technologies5030050}
\showDOI{\tempurl}


\bibitem[\protect\citeauthoryear{Ashbrook, Guo, and Lambie}{Ashbrook
  et~al\mbox{.}}{2016}]%
        {ashbrookAugmentedFabricationCombining2016}
\bibfield{author}{\bibinfo{person}{Daniel Ashbrook},
  \bibinfo{person}{Shitao~Stan Guo}, {and} \bibinfo{person}{Alan Lambie}.}
  \bibinfo{year}{2016}\natexlab{}.
\newblock \showarticletitle{Towards {{Augmented Fabrication}}: Combining
  {{Fabricated}} and {{Existing Objects}}}. In
  \bibinfo{booktitle}{\emph{Proceedings of the 2016 {{CHI Conference Extended
  Abstracts}} on {{Human Factors}} in {{Computing Systems}}}}
  \emph{(\bibinfo{series}{{{CHI EA}} '16})}. \bibinfo{publisher}{{ACM}},
  \bibinfo{address}{{New York, NY, USA}}, \bibinfo{pages}{1510--1518}.
\newblock
\showISBNx{978-1-4503-4082-3}
\urldef\tempurl%
\url{https://doi.org/10.1145/2851581.2892509}
\showDOI{\tempurl}


\bibitem[\protect\citeauthoryear{Auger}{Auger}{2013}]%
        {augerSpeculativeDesignCrafting2013}
\bibfield{author}{\bibinfo{person}{James Auger}.}
  \bibinfo{year}{2013}\natexlab{}.
\newblock \showarticletitle{Speculative Design: Crafting the Speculation}.
\newblock \bibinfo{journal}{\emph{Digital Creativity}} \bibinfo{volume}{24},
  \bibinfo{number}{1} (\bibinfo{date}{March} \bibinfo{year}{2013}),
  \bibinfo{pages}{11--35}.
\newblock
\showISSN{1462-6268}
\urldef\tempurl%
\url{https://doi.org/10.1080/14626268.2013.767276}
\showDOI{\tempurl}


\bibitem[\protect\citeauthoryear{Baker, Miner, and Eesley}{Baker
  et~al\mbox{.}}{2003}]%
        {bakerImprovisingFirmsBricolage2003}
\bibfield{author}{\bibinfo{person}{Ted Baker}, \bibinfo{person}{Anne~S. Miner},
  {and} \bibinfo{person}{Dale~T. Eesley}.} \bibinfo{year}{2003}\natexlab{}.
\newblock \showarticletitle{Improvising Firms: Bricolage, Account Giving and
  Improvisational Competencies in the Founding Process}.
\newblock \bibinfo{journal}{\emph{Research policy}} \bibinfo{volume}{32},
  \bibinfo{number}{2} (\bibinfo{year}{2003}), \bibinfo{pages}{255--276}.
\newblock


\bibitem[\protect\citeauthoryear{Barni, Carpanzano, Landolfi, and
  Pedrazzoli}{Barni et~al\mbox{.}}{2019}]%
        {barniUrbanManufacturingSustainable2019}
\bibfield{author}{\bibinfo{person}{Andrea Barni}, \bibinfo{person}{Emanuele
  Carpanzano}, \bibinfo{person}{Giuseppe Landolfi}, {and}
  \bibinfo{person}{Paolo Pedrazzoli}.} \bibinfo{year}{2019}\natexlab{}.
\newblock \showarticletitle{Urban {{Manufacturing}} of {{Sustainable
  Customer}}-{{Oriented Products}}}.
\newblock In \bibinfo{booktitle}{\emph{Plant {{Innate Immunity}}}},
  \bibfield{editor}{\bibinfo{person}{Walter Gassmann}} (Ed.).
  Vol.~\bibinfo{volume}{1991}. \bibinfo{publisher}{{Springer New York}},
  \bibinfo{address}{{New York, NY}}, \bibinfo{pages}{128--141}.
\newblock
\showISBNx{978-1-4939-9457-1 978-1-4939-9458-8}
\urldef\tempurl%
\url{https://doi.org/10.1007/978-3-030-18180-2_10}
\showDOI{\tempurl}


\bibitem[\protect\citeauthoryear{Baudisch and Mueller}{Baudisch and
  Mueller}{2017}]%
        {baudischPersonalFabrication2017}
\bibfield{author}{\bibinfo{person}{Patrick Baudisch} {and}
  \bibinfo{person}{Stefanie Mueller}.} \bibinfo{year}{2017}\natexlab{}.
\newblock \showarticletitle{Personal {{Fabrication}}}.
\newblock \bibinfo{journal}{\emph{Foundations and Trends\textregistered{} in
  Human\textendash Computer Interaction}} \bibinfo{volume}{10},
  \bibinfo{number}{3\textendash 4} (\bibinfo{date}{May} \bibinfo{year}{2017}),
  \bibinfo{pages}{165--293}.
\newblock
\showISSN{1551-3955, 1551-3963}
\urldef\tempurl%
\url{https://doi.org/10.1561/1100000055}
\showDOI{\tempurl}


\bibitem[\protect\citeauthoryear{Bellows}{Bellows}{1936}]%
        {bellowsPoeticEdda1936}
\bibfield{author}{\bibinfo{person}{Henry~Adams Bellows}.}
  \bibinfo{year}{1936}\natexlab{}.
\newblock \bibinfo{booktitle}{\emph{The Poetic Edda}}.
  Vol.~\bibinfo{volume}{1}.
\newblock \bibinfo{publisher}{{Princeton University Press Princeton, NJ, USA}},
  \bibinfo{address}{{Princeton, NJ, USA}}.
\newblock


\bibitem[\protect\citeauthoryear{Berman, Quek, Woodward, Okundaye, and
  Kim}{Berman et~al\mbox{.}}{2020}]%
        {bermanAnyoneCanPrint2020}
\bibfield{author}{\bibinfo{person}{Alexander Berman}, \bibinfo{person}{Francis
  Quek}, \bibinfo{person}{Robert Woodward}, \bibinfo{person}{Osazuwa Okundaye},
  {and} \bibinfo{person}{Jeeeun Kim}.} \bibinfo{year}{2020}\natexlab{}.
\newblock \showarticletitle{\&\#x201c;{{Anyone Can Print}}\&\#x201d;:
  Supporting {{Collaborations}} with {{3D Printing Services}} to {{Empower
  Broader Participation}} in {{Personal Fabrication}}}. In
  \bibinfo{booktitle}{\emph{Proceedings of the 11th {{Nordic Conference}} on
  {{Human}}-{{Computer Interaction}}: Shaping {{Experiences}}, {{Shaping
  Society}}}} \emph{(\bibinfo{series}{{{NordiCHI}} '20})}.
  \bibinfo{publisher}{{Association for Computing Machinery}},
  \bibinfo{address}{{New York, NY, USA}}, \bibinfo{pages}{1--13}.
\newblock
\showISBNx{978-1-4503-7579-5}
\urldef\tempurl%
\url{https://doi.org/10.1145/3419249.3420068}
\showDOI{\tempurl}


\bibitem[\protect\citeauthoryear{Berman, Thakare, Howell, Quek, and Kim}{Berman
  et~al\mbox{.}}{2021}]%
        {bermanHowDIYMetaDesignTools2021}
\bibfield{author}{\bibinfo{person}{Alexander Berman}, \bibinfo{person}{Ketan
  Thakare}, \bibinfo{person}{Joshua Howell}, \bibinfo{person}{Francis Quek},
  {and} \bibinfo{person}{Jeeeun Kim}.} \bibinfo{year}{2021}\natexlab{}.
\newblock \showarticletitle{{{HowDIY}}: Towards {{Meta}}-{{Design Tools}} to
  {{Support Anyone}} to {{3D Print Anywhere}}}. In
  \bibinfo{booktitle}{\emph{26th {{International Conference}} on {{Intelligent
  User Interfaces}}}} \emph{(\bibinfo{series}{{{IUI}} '21})}.
  \bibinfo{publisher}{{Association for Computing Machinery}},
  \bibinfo{address}{{New York, NY, USA}}, \bibinfo{pages}{491--503}.
\newblock
\showISBNx{978-1-4503-8017-1}
\urldef\tempurl%
\url{https://doi.org/10.1145/3397481.3450638}
\showDOI{\tempurl}


\bibitem[\protect\citeauthoryear{Blevis}{Blevis}{2007}]%
        {blevisSustainableInteractionDesign2007}
\bibfield{author}{\bibinfo{person}{Eli Blevis}.}
  \bibinfo{year}{2007}\natexlab{}.
\newblock \showarticletitle{Sustainable {{Interaction Design}}: Invention \&
  {{Disposal}}, {{Renewal}} \& {{Reuse}}}. In
  \bibinfo{booktitle}{\emph{Proceedings of the {{SIGCHI Conference}} on {{Human
  Factors}} in {{Computing Systems}}}} \emph{(\bibinfo{series}{{{CHI}} '07})}.
  \bibinfo{publisher}{{ACM}}, \bibinfo{address}{{New York, NY, USA}},
  \bibinfo{pages}{503--512}.
\newblock
\showISBNx{978-1-59593-593-9}
\urldef\tempurl%
\url{https://doi.org/10.1145/1240624.1240705}
\showDOI{\tempurl}


\bibitem[\protect\citeauthoryear{Bosch}{Bosch}{2012}]%
        {boschSciFiWriterBruce2012}
\bibfield{author}{\bibinfo{person}{Torie Bosch}.}
  \bibinfo{year}{2012}\natexlab{}.
\newblock \bibinfo{title}{Sci-{{Fi Writer Bruce Sterling Explains}} the
  {{Intriguing New Concept}} of {{Design Fiction}}}.
\newblock
  \bibinfo{howpublished}{https://slate.com/technology/2012/03/bruce-sterling-on-design-fictions.html}.
\newblock
\newblock
\shownote{(Accessed 25.04.2021)}.


\bibitem[\protect\citeauthoryear{Brand}{Brand}{1995}]%
        {brandHowBuildingsLearn1995}
\bibfield{author}{\bibinfo{person}{Stewart Brand}.}
  \bibinfo{year}{1995}\natexlab{}.
\newblock \bibinfo{booktitle}{\emph{How {{Buildings Learn}}: What {{Happens
  After They}}'re {{Built}}}}.
\newblock \bibinfo{publisher}{{Penguin}}, \bibinfo{address}{{City of
  Westminster, London, England}}.
\newblock
\showISBNx{978-1-101-56264-2}


\bibitem[\protect\citeauthoryear{Brockman}{Brockman}{2013}]%
        {brockmanThisExplainsEverything2013}
\bibfield{author}{\bibinfo{person}{John Brockman}.}
  \bibinfo{year}{2013}\natexlab{}.
\newblock \bibinfo{booktitle}{\emph{This {{Explains Everything}}: Deep,
  {{Beautiful}}, and {{Elegant Theories}} of {{How}} the {{World Works}}}}.
\newblock \bibinfo{publisher}{{Harper Perennial}}, \bibinfo{address}{{New York,
  United States}}.
\newblock
\showISBNx{978-0-06-223017-1}


\bibitem[\protect\citeauthoryear{Chen, Coros, Mankoff, and Hudson}{Chen
  et~al\mbox{.}}{2015}]%
        {chenEncore3DPrinted2015}
\bibfield{author}{\bibinfo{person}{Xiang~'Anthony' Chen},
  \bibinfo{person}{Stelian Coros}, \bibinfo{person}{Jennifer Mankoff}, {and}
  \bibinfo{person}{Scott~E. Hudson}.} \bibinfo{year}{2015}\natexlab{}.
\newblock \showarticletitle{Encore: {{3D Printed Augmentation}} of {{Everyday
  Objects}} with {{Printed}}-{{Over}}, {{Affixed}} and {{Interlocked
  Attachments}}}. In \bibinfo{booktitle}{\emph{Proceedings of the 28th {{Annual
  ACM Symposium}} on {{User Interface Software}} \& {{Technology}} - {{UIST}}
  '15}}. \bibinfo{publisher}{{ACM Press}}, \bibinfo{address}{{Daegu, Kyungpook,
  Republic of Korea}}, \bibinfo{pages}{73--82}.
\newblock
\showISBNx{978-1-4503-3779-3}
\urldef\tempurl%
\url{https://doi.org/10.1145/2807442.2807498}
\showDOI{\tempurl}


\bibitem[\protect\citeauthoryear{Chen, Kim, Mankoff, Grossman, Coros, and
  Hudson}{Chen et~al\mbox{.}}{2016}]%
        {chenRepriseDesignTool2016}
\bibfield{author}{\bibinfo{person}{Xiang~'Anthony' Chen},
  \bibinfo{person}{Jeeeun Kim}, \bibinfo{person}{Jennifer Mankoff},
  \bibinfo{person}{Tovi Grossman}, \bibinfo{person}{Stelian Coros}, {and}
  \bibinfo{person}{Scott~E. Hudson}.} \bibinfo{year}{2016}\natexlab{}.
\newblock \showarticletitle{Reprise: A {{Design Tool}} for {{Specifying}},
  {{Generating}}, and {{Customizing 3D Printable Adaptations}} on {{Everyday
  Objects}}}. In \bibinfo{booktitle}{\emph{Proceedings of the 29th {{Annual
  Symposium}} on {{User Interface Software}} and {{Technology}}}}
  \emph{(\bibinfo{series}{{{UIST}} '16})}. \bibinfo{publisher}{{ACM}},
  \bibinfo{address}{{New York, NY, USA}}, \bibinfo{pages}{29--39}.
\newblock
\showISBNx{978-1-4503-4189-9}
\urldef\tempurl%
\url{https://doi.org/10.1145/2984511.2984512}
\showDOI{\tempurl}


\bibitem[\protect\citeauthoryear{Choi and Ishii}{Choi and Ishii}{2021}]%
        {choiThermsUpDIYInflatables2021}
\bibfield{author}{\bibinfo{person}{Kyung~Yun Choi} {and}
  \bibinfo{person}{Hiroshi Ishii}.} \bibinfo{year}{2021}\natexlab{}.
\newblock \showarticletitle{Therms-{{Up}}!: {{DIY Inflatables}} and
  {{Interactive Materials}} by {{Upcycling Wasted Thermoplastic Bags}}}. In
  \bibinfo{booktitle}{\emph{Proceedings of the {{Fifteenth International
  Conference}} on {{Tangible}}, {{Embedded}}, and {{Embodied Interaction}}}}
  \emph{(\bibinfo{series}{{{TEI}} '21})}. \bibinfo{publisher}{{Association for
  Computing Machinery}}, \bibinfo{address}{{New York, NY, USA}},
  \bibinfo{pages}{1--8}.
\newblock
\showISBNx{978-1-4503-8213-7}
\urldef\tempurl%
\url{https://doi.org/10.1145/3430524.3442457}
\showDOI{\tempurl}


\bibitem[\protect\citeauthoryear{Collingridge}{Collingridge}{1982}]%
        {collingridgeSocialControlTechnology1982}
\bibfield{author}{\bibinfo{person}{David Collingridge}.}
  \bibinfo{year}{1982}\natexlab{}.
\newblock \bibinfo{booktitle}{\emph{The {{Social Control}} of {{Technology}}}
  (\bibinfo{edition}{repr} ed.)}.
\newblock \bibinfo{publisher}{{Pinter [u.a.]}}, \bibinfo{address}{{London}}.
\newblock
\showISBNx{978-0-903804-72-1 978-0-312-73168-7}


\bibitem[\protect\citeauthoryear{Devendorf and Ryokai}{Devendorf and
  Ryokai}{2015}]%
        {devendorfBeingMachineReconfiguring2015}
\bibfield{author}{\bibinfo{person}{Laura Devendorf} {and}
  \bibinfo{person}{Kimiko Ryokai}.} \bibinfo{year}{2015}\natexlab{}.
\newblock \showarticletitle{Being the {{Machine}}: Reconfiguring {{Agency}} and
  {{Control}} in {{Hybrid Fabrication}}}. In
  \bibinfo{booktitle}{\emph{Proceedings of the 33rd {{Annual ACM Conference}}
  on {{Human Factors}} in {{Computing Systems}}}}
  \emph{(\bibinfo{series}{{{CHI}} '15})}. \bibinfo{publisher}{{ACM}},
  \bibinfo{address}{{New York, NY, USA}}, \bibinfo{pages}{2477--2486}.
\newblock
\showISBNx{978-1-4503-3145-6}
\urldef\tempurl%
\url{https://doi.org/10.1145/2702123.2702547}
\showDOI{\tempurl}


\bibitem[\protect\citeauthoryear{Dew and Rosner}{Dew and Rosner}{2019}]%
        {dewDesigningWasteSituated2019}
\bibfield{author}{\bibinfo{person}{Kristin~N. Dew} {and}
  \bibinfo{person}{Daniela~K. Rosner}.} \bibinfo{year}{2019}\natexlab{}.
\newblock \showarticletitle{Designing with {{Waste}}: A {{Situated Inquiry}}
  into the {{Material Excess}} of {{Making}}}. In
  \bibinfo{booktitle}{\emph{Proceedings of the 2019 on {{Designing Interactive
  Systems Conference}}}} \emph{(\bibinfo{series}{{{DIS}} '19})}.
  \bibinfo{publisher}{{Association for Computing Machinery}},
  \bibinfo{address}{{New York, NY, USA}}, \bibinfo{pages}{1307--1319}.
\newblock
\showISBNx{978-1-4503-5850-7}
\urldef\tempurl%
\url{https://doi.org/10.1145/3322276.3322320}
\showDOI{\tempurl}


\bibitem[\protect\citeauthoryear{D{\"o}ring, Sylvester, and Schmidt}{D{\"o}ring
  et~al\mbox{.}}{2013}]%
        {doringDesignSpaceEphemeral2013}
\bibfield{author}{\bibinfo{person}{Tanja D{\"o}ring}, \bibinfo{person}{Axel
  Sylvester}, {and} \bibinfo{person}{Albrecht Schmidt}.}
  \bibinfo{year}{2013}\natexlab{}.
\newblock \showarticletitle{A {{Design Space}} for {{Ephemeral User
  Interfaces}}}. In \bibinfo{booktitle}{\emph{Proceedings of the 7th
  {{International Conference}} on {{Tangible}}, {{Embedded}} and {{Embodied
  Interaction}} - {{TEI}} '13}}. \bibinfo{publisher}{{ACM Press}},
  \bibinfo{address}{{Barcelona, Spain}}, \bibinfo{pages}{75}.
\newblock
\showISBNx{978-1-4503-1898-3}
\urldef\tempurl%
\url{https://doi.org/10.1145/2460625.2460637}
\showDOI{\tempurl}


\bibitem[\protect\citeauthoryear{Dunne and Raby}{Dunne and Raby}{2013}]%
        {dunneSpeculativeEverythingDesign2013}
\bibfield{author}{\bibinfo{person}{Anthony Dunne} {and} \bibinfo{person}{Fiona
  Raby}.} \bibinfo{year}{2013}\natexlab{}.
\newblock \bibinfo{booktitle}{\emph{Speculative {{Everything}}: Design,
  {{Fiction}}, and {{Social Dreaming}}}}.
\newblock \bibinfo{publisher}{{MIT Press}}, \bibinfo{address}{{Cambridge, MA,
  USA}}.
\newblock
\showISBNx{978-0-262-01984-2}


\bibitem[\protect\citeauthoryear{Duymedjian and R{\"u}ling}{Duymedjian and
  R{\"u}ling}{2010}]%
        {duymedjianFoundationBricolageOrganization2010}
\bibfield{author}{\bibinfo{person}{Raffi Duymedjian} {and}
  \bibinfo{person}{Charles-Clemens R{\"u}ling}.}
  \bibinfo{year}{2010}\natexlab{}.
\newblock \showarticletitle{Towards a Foundation of Bricolage in Organization
  and Management Theory}.
\newblock \bibinfo{journal}{\emph{Organization Studies}} \bibinfo{volume}{31},
  \bibinfo{number}{2} (\bibinfo{year}{2010}), \bibinfo{pages}{133--151}.
\newblock


\bibitem[\protect\citeauthoryear{Efrat, Mizrahi, and Zoran}{Efrat
  et~al\mbox{.}}{2016}]%
        {efratHybridBricolageBridging2016}
\bibfield{author}{\bibinfo{person}{Tamara~Anna Efrat}, \bibinfo{person}{Moran
  Mizrahi}, {and} \bibinfo{person}{Amit Zoran}.}
  \bibinfo{year}{2016}\natexlab{}.
\newblock \showarticletitle{The {{Hybrid Bricolage}}: Bridging {{Parametric
  Design}} with {{Craft Through Algorithmic Modularity}}}. In
  \bibinfo{booktitle}{\emph{Proceedings of the 2016 {{CHI Conference}} on
  {{Human Factors}} in {{Computing Systems}}}} \emph{(\bibinfo{series}{{{CHI}}
  '16})}. \bibinfo{publisher}{{ACM}}, \bibinfo{address}{{New York, NY, USA}},
  \bibinfo{pages}{5984--5995}.
\newblock
\showISBNx{978-1-4503-3362-7}
\urldef\tempurl%
\url{https://doi.org/10.1145/2858036.2858441}
\showDOI{\tempurl}


\bibitem[\protect\citeauthoryear{Fernandes}{Fernandes}{2016}]%
        {fernandesHumanAugmentationWearables2016}
\bibfield{author}{\bibinfo{person}{Tony Fernandes}.}
  \bibinfo{year}{2016}\natexlab{}.
\newblock \showarticletitle{Human Augmentation: Beyond Wearables}.
\newblock \bibinfo{journal}{\emph{Interactions}} \bibinfo{volume}{23},
  \bibinfo{number}{5} (\bibinfo{date}{Aug.} \bibinfo{year}{2016}),
  \bibinfo{pages}{66--68}.
\newblock
\showISSN{1072-5520}
\urldef\tempurl%
\url{https://doi.org/10.1145/2972228}
\showDOI{\tempurl}


\bibitem[\protect\citeauthoryear{Gannon, Grossman, and Fitzmaurice}{Gannon
  et~al\mbox{.}}{2015}]%
        {gannonTactumSkinCentricApproach2015}
\bibfield{author}{\bibinfo{person}{Madeline Gannon}, \bibinfo{person}{Tovi
  Grossman}, {and} \bibinfo{person}{George Fitzmaurice}.}
  \bibinfo{year}{2015}\natexlab{}.
\newblock \showarticletitle{Tactum: A {{Skin}}-{{Centric Approach}} to
  {{Digital Design}} and {{Fabrication}}}. In
  \bibinfo{booktitle}{\emph{Proceedings of the 33rd {{Annual ACM Conference}}
  on {{Human Factors}} in {{Computing Systems}}}}
  \emph{(\bibinfo{series}{{{CHI}} '15})}. \bibinfo{publisher}{{Association for
  Computing Machinery}}, \bibinfo{address}{{Seoul, Republic of Korea}},
  \bibinfo{pages}{1779--1788}.
\newblock
\showISBNx{978-1-4503-3145-6}
\urldef\tempurl%
\url{https://doi.org/10.1145/2702123.2702581}
\showDOI{\tempurl}


\bibitem[\protect\citeauthoryear{Gannon, Grossman, and Fitzmaurice}{Gannon
  et~al\mbox{.}}{2016}]%
        {gannonExoSkinOnBodyFabrication2016}
\bibfield{author}{\bibinfo{person}{Madeline Gannon}, \bibinfo{person}{Tovi
  Grossman}, {and} \bibinfo{person}{George Fitzmaurice}.}
  \bibinfo{year}{2016}\natexlab{}.
\newblock \showarticletitle{{{ExoSkin}}: On-{{Body Fabrication}}}. In
  \bibinfo{booktitle}{\emph{Proceedings of the 2016 {{CHI Conference}} on
  {{Human Factors}} in {{Computing Systems}}}} \emph{(\bibinfo{series}{{{CHI}}
  '16})}. \bibinfo{publisher}{{ACM}}, \bibinfo{address}{{New York, NY, USA}},
  \bibinfo{pages}{5996--6007}.
\newblock
\showISBNx{978-1-4503-3362-7}
\urldef\tempurl%
\url{https://doi.org/10.1145/2858036.2858576}
\showDOI{\tempurl}


\bibitem[\protect\citeauthoryear{Gershenfeld, Gershenfeld, and
  {Cutcher-Gershenfeld}}{Gershenfeld et~al\mbox{.}}{2017}]%
        {gershenfeldDesigningRealityHow2017}
\bibfield{author}{\bibinfo{person}{Neil Gershenfeld}, \bibinfo{person}{Alan
  Gershenfeld}, {and} \bibinfo{person}{Joel {Cutcher-Gershenfeld}}.}
  \bibinfo{year}{2017}\natexlab{}.
\newblock \bibinfo{booktitle}{\emph{Designing {{Reality}}: How to {{Survive}}
  and {{Thrive}} in the {{Third Digital Revolution}}}}.
\newblock \bibinfo{publisher}{{Basic Books, Inc.}}, \bibinfo{address}{{USA}}.
\newblock
\showISBNx{978-0-465-09347-2}


\bibitem[\protect\citeauthoryear{Gonz{\'a}lez, Salgado, Garc{\'i}a~Moruno, and
  S{\'a}nchez~R{\'i}os}{Gonz{\'a}lez et~al\mbox{.}}{2018}]%
        {gonzalezErgonomicCustomizedToolHandle2018}
\bibfield{author}{\bibinfo{person}{Alfonso Gonz{\'a}lez},
  \bibinfo{person}{David Salgado}, \bibinfo{person}{Lorenzo Garc{\'i}a~Moruno},
  {and} \bibinfo{person}{Alonso S{\'a}nchez~R{\'i}os}.}
  \bibinfo{year}{2018}\natexlab{}.
\newblock \showarticletitle{An {{Ergonomic Customized}}-{{Tool Handle Design}}
  for {{Precision Tools}} Using {{Additive Manufacturing}}: A {{Case Study}}}.
\newblock \bibinfo{journal}{\emph{Applied Sciences}} \bibinfo{volume}{8},
  \bibinfo{number}{7} (\bibinfo{date}{July} \bibinfo{year}{2018}),
  \bibinfo{pages}{1200}.
\newblock
\showISSN{2076-3417}
\urldef\tempurl%
\url{https://doi.org/10.3390/app8071200}
\showDOI{\tempurl}


\bibitem[\protect\citeauthoryear{Hu, Leigh, and Maes}{Hu et~al\mbox{.}}{2017}]%
        {huHandDevelopmentKit2017}
\bibfield{author}{\bibinfo{person}{Yuhan Hu}, \bibinfo{person}{Sang-won Leigh},
  {and} \bibinfo{person}{Pattie Maes}.} \bibinfo{year}{2017}\natexlab{}.
\newblock \showarticletitle{Hand {{Development Kit}}: Soft {{Robotic Fingers As
  Prosthetic Augmentation}} of the {{Hand}}}. In
  \bibinfo{booktitle}{\emph{Adjunct {{Publication}} of the 30th {{Annual ACM
  Symposium}} on {{User Interface Software}} and {{Technology}}}}
  \emph{(\bibinfo{series}{{{UIST}} '17})}. \bibinfo{publisher}{{ACM}},
  \bibinfo{address}{{New York, NY, USA}}, \bibinfo{pages}{27--29}.
\newblock
\showISBNx{978-1-4503-5419-6}
\urldef\tempurl%
\url{https://doi.org/10.1145/3131785.3131805}
\showDOI{\tempurl}


\bibitem[\protect\citeauthoryear{Ishii, Lakatos, Bonanni, and Labrune}{Ishii
  et~al\mbox{.}}{2012}]%
        {ishiiRadicalAtomsTangible2012}
\bibfield{author}{\bibinfo{person}{Hiroshi Ishii}, \bibinfo{person}{D{\'a}vid
  Lakatos}, \bibinfo{person}{Leonardo Bonanni}, {and}
  \bibinfo{person}{Jean-Baptiste Labrune}.} \bibinfo{year}{2012}\natexlab{}.
\newblock \showarticletitle{Radical {{Atoms}}: Beyond {{Tangible Bits}},
  {{Toward Transformable Materials}}}.
\newblock \bibinfo{journal}{\emph{interactions}} \bibinfo{volume}{19},
  \bibinfo{number}{1} (\bibinfo{date}{Jan.} \bibinfo{year}{2012}),
  \bibinfo{pages}{38--51}.
\newblock
\showISSN{1072-5520}
\urldef\tempurl%
\url{https://doi.org/10.1145/2065327.2065337}
\showDOI{\tempurl}


\bibitem[\protect\citeauthoryear{Jackson}{Jackson}{2017}]%
        {jacksonHowAnimalsFabricate2017}
\bibfield{author}{\bibinfo{person}{Daniel~John Jackson}.}
  \bibinfo{year}{2017}\natexlab{}.
\newblock \showarticletitle{How {{Animals Fabricate Biominerals}}}.
\newblock \bibinfo{journal}{\emph{Scientia}}  \bibinfo{volume}{Biology, Earth
  and Environment.} (\bibinfo{date}{Feb.} \bibinfo{year}{2017}),
  \bibinfo{pages}{3}.
\newblock


\bibitem[\protect\citeauthoryear{Johnson}{Johnson}{2012}]%
        {johnsonBricoleurBricolageMetaphor2012}
\bibfield{author}{\bibinfo{person}{Christopher Johnson}.}
  \bibinfo{year}{2012}\natexlab{}.
\newblock \showarticletitle{Bricoleur and Bricolage: From Metaphor to Universal
  Concept}.
\newblock \bibinfo{journal}{\emph{Paragraph}} \bibinfo{volume}{35},
  \bibinfo{number}{3} (\bibinfo{year}{2012}), \bibinfo{pages}{355--372}.
\newblock


\bibitem[\protect\citeauthoryear{Jones, Seppi, and Olsen}{Jones
  et~al\mbox{.}}{2016}]%
        {jonesWhatYouSculpt2016}
\bibfield{author}{\bibinfo{person}{Michael~D. Jones}, \bibinfo{person}{Kevin
  Seppi}, {and} \bibinfo{person}{Dan~R. Olsen}.}
  \bibinfo{year}{2016}\natexlab{}.
\newblock \showarticletitle{What {{You Sculpt}} Is {{What You Get}}: Modeling
  {{Physical Interactive Devices}} with {{Clay}} and {{3D Printed Widgets}}}.
  In \bibinfo{booktitle}{\emph{Proceedings of the 2016 {{CHI Conference}} on
  {{Human Factors}} in {{Computing Systems}}}} \emph{(\bibinfo{series}{{{CHI}}
  '16})}. \bibinfo{publisher}{{ACM}}, \bibinfo{address}{{New York, NY, USA}},
  \bibinfo{pages}{876--886}.
\newblock
\showISBNx{978-1-4503-3362-7}
\urldef\tempurl%
\url{https://doi.org/10.1145/2858036.2858493}
\showDOI{\tempurl}


\bibitem[\protect\citeauthoryear{Ko, Yim, Lee, Pyun, and Lee}{Ko
  et~al\mbox{.}}{2021}]%
        {koDesigningMetamaterialCells2021}
\bibfield{author}{\bibinfo{person}{Donghyeon Ko}, \bibinfo{person}{Jee~Bin
  Yim}, \bibinfo{person}{Yujin Lee}, \bibinfo{person}{Jaehoon Pyun}, {and}
  \bibinfo{person}{Woohun Lee}.} \bibinfo{year}{2021}\natexlab{}.
\newblock \showarticletitle{Designing {{Metamaterial Cells}} to {{Enrich
  Thermoforming 3D Printed Object}} for {{Post}}-{{Print Modification}}}. In
  \bibinfo{booktitle}{\emph{Proceedings of the 2021 {{CHI Conference}} on
  {{Human Factors}} in {{Computing Systems}}}} \emph{(\bibinfo{series}{{{CHI}}
  '21})}. \bibinfo{publisher}{{Association for Computing Machinery}},
  \bibinfo{address}{{New York, NY, USA}}, \bibinfo{pages}{1--12}.
\newblock
\showISBNx{978-1-4503-8096-6}
\urldef\tempurl%
\url{https://doi.org/10.1145/3411764.3445229}
\showDOI{\tempurl}


\bibitem[\protect\citeauthoryear{Kohtala}{Kohtala}{2017}]%
        {kohtalaMakingMakingCritical2017}
\bibfield{author}{\bibinfo{person}{Cindy Kohtala}.}
  \bibinfo{year}{2017}\natexlab{}.
\newblock \showarticletitle{Making ``{{Making}}'' {{Critical}}: How
  {{Sustainability}} Is {{Constituted}} in {{Fab Lab Ideology}}}.
\newblock \bibinfo{journal}{\emph{The Design Journal}} \bibinfo{volume}{20},
  \bibinfo{number}{3} (\bibinfo{date}{May} \bibinfo{year}{2017}),
  \bibinfo{pages}{375--394}.
\newblock
\showISSN{1460-6925}
\urldef\tempurl%
\url{https://doi.org/10.1080/14606925.2016.1261504}
\showDOI{\tempurl}


\bibitem[\protect\citeauthoryear{Kwon, Jaiswal, Benford, Seah, Bennett, Koleva,
  and Schn{\"a}delbach}{Kwon et~al\mbox{.}}{2015}]%
        {kwonFugaciousFilmExploringAttentive2015}
\bibfield{author}{\bibinfo{person}{Hyosun Kwon}, \bibinfo{person}{Shashank
  Jaiswal}, \bibinfo{person}{Steve Benford}, \bibinfo{person}{Sue~Ann Seah},
  \bibinfo{person}{Peter Bennett}, \bibinfo{person}{Boriana Koleva}, {and}
  \bibinfo{person}{Holger Schn{\"a}delbach}.} \bibinfo{year}{2015}\natexlab{}.
\newblock \showarticletitle{{{FugaciousFilm}}: Exploring {{Attentive
  Interaction}} with {{Ephemeral Material}}}. In
  \bibinfo{booktitle}{\emph{Proceedings of the 33rd {{Annual ACM Conference}}
  on {{Human Factors}} in {{Computing Systems}}}}
  \emph{(\bibinfo{series}{{{CHI}} '15})}. \bibinfo{publisher}{{ACM}},
  \bibinfo{address}{{New York, NY, USA}}, \bibinfo{pages}{1285--1294}.
\newblock
\showISBNx{978-1-4503-3145-6}
\urldef\tempurl%
\url{https://doi.org/10.1145/2702123.2702206}
\showDOI{\tempurl}


\bibitem[\protect\citeauthoryear{Leen, Ramakers, and Luyten}{Leen
  et~al\mbox{.}}{2017}]%
        {leenStrutModelingLowFidelityConstruction2017}
\bibfield{author}{\bibinfo{person}{Danny Leen}, \bibinfo{person}{Raf Ramakers},
  {and} \bibinfo{person}{Kris Luyten}.} \bibinfo{year}{2017}\natexlab{}.
\newblock \showarticletitle{{{StrutModeling}}: A {{Low}}-{{Fidelity
  Construction Kit}} to {{Iteratively Model}}, {{Test}}, and {{Adapt 3D
  Objects}}}. In \bibinfo{booktitle}{\emph{Proceedings of the 30th {{Annual ACM
  Symposium}} on {{User Interface Software}} and {{Technology}}}}
  \emph{(\bibinfo{series}{{{UIST}} '17})}. \bibinfo{publisher}{{ACM}},
  \bibinfo{address}{{New York, NY, USA}}, \bibinfo{pages}{471--479}.
\newblock
\showISBNx{978-1-4503-4981-9}
\urldef\tempurl%
\url{https://doi.org/10.1145/3126594.3126643}
\showDOI{\tempurl}


\bibitem[\protect\citeauthoryear{Leigh, Denton, Parekh, Peebles, Johnson, and
  Maes}{Leigh et~al\mbox{.}}{2018}]%
        {leighMorphologyExtensionKit2018}
\bibfield{author}{\bibinfo{person}{Sang-won Leigh}, \bibinfo{person}{Timothy
  Denton}, \bibinfo{person}{Kush Parekh}, \bibinfo{person}{William Peebles},
  \bibinfo{person}{Magnus Johnson}, {and} \bibinfo{person}{Pattie Maes}.}
  \bibinfo{year}{2018}\natexlab{}.
\newblock \showarticletitle{Morphology {{Extension Kit}}: A {{Modular Robotic
  Platform}} for {{Physically Reconfigurable Wearables}}}. In
  \bibinfo{booktitle}{\emph{Proceedings of the {{Twelfth International
  Conference}} on {{Tangible}}, {{Embedded}}, and {{Embodied Interaction}}}}
  \emph{(\bibinfo{series}{{{TEI}} '18})}. \bibinfo{publisher}{{ACM}},
  \bibinfo{address}{{New York, NY, USA}}, \bibinfo{pages}{11--18}.
\newblock
\showISBNx{978-1-4503-5568-1}
\urldef\tempurl%
\url{https://doi.org/10.1145/3173225.3173239}
\showDOI{\tempurl}


\bibitem[\protect\citeauthoryear{Leigh and Maes}{Leigh and Maes}{2016}]%
        {leighBodyIntegratedProgrammable2016}
\bibfield{author}{\bibinfo{person}{Sang-won Leigh} {and}
  \bibinfo{person}{Pattie Maes}.} \bibinfo{year}{2016}\natexlab{}.
\newblock \showarticletitle{Body {{Integrated Programmable Joints Interface}}}.
  In \bibinfo{booktitle}{\emph{Proceedings of the 2016 {{CHI Conference}} on
  {{Human Factors}} in {{Computing Systems}}}}. \bibinfo{publisher}{{ACM}},
  \bibinfo{address}{{New York, NY, USA}}, \bibinfo{pages}{6053--6057}.
\newblock


\bibitem[\protect\citeauthoryear{{Levi-Strauss}}{{Levi-Strauss}}{1966}]%
        {levi-straussSavageMindPensee1966}
\bibfield{author}{\bibinfo{person}{Claude {Levi-Strauss}}.}
  \bibinfo{year}{1966}\natexlab{}.
\newblock \bibinfo{booktitle}{\emph{The {{Savage Mind}} ({{La Pens\'ee
  Sauvage}})}}.
\newblock \bibinfo{publisher}{{Weidenfeld and Nicolson}},
  \bibinfo{address}{{Librairie Plon, 8, rue Garanciere, Paris-6\textbullet}}.
\newblock


\bibitem[\protect\citeauthoryear{Lewis and Narayan}{Lewis and Narayan}{1993}]%
        {lewisDesignSizingErgonomic1993}
\bibfield{author}{\bibinfo{person}{Winston~G. Lewis} {and}
  \bibinfo{person}{C.~V. Narayan}.} \bibinfo{year}{1993}\natexlab{}.
\newblock \showarticletitle{Design and Sizing of Ergonomic Handles for Hand
  Tools}.
\newblock \bibinfo{journal}{\emph{Applied Ergonomics}} \bibinfo{volume}{24},
  \bibinfo{number}{5} (\bibinfo{date}{Oct.} \bibinfo{year}{1993}),
  \bibinfo{pages}{351--356}.
\newblock
\showISSN{0003-6870}
\urldef\tempurl%
\url{https://doi.org/10.1016/0003-6870(93)90074-J}
\showDOI{\tempurl}


\bibitem[\protect\citeauthoryear{Licklider}{Licklider}{1960}]%
        {lickliderManComputerSymbiosis1960}
\bibfield{author}{\bibinfo{person}{J.~C.~R. Licklider}.}
  \bibinfo{year}{1960}\natexlab{}.
\newblock \showarticletitle{Man-{{Computer Symbiosis}}}.
\newblock \bibinfo{journal}{\emph{IRE Transactions on Human Factors in
  Electronics}} \bibinfo{volume}{HFE-1}, \bibinfo{number}{1}
  (\bibinfo{date}{March} \bibinfo{year}{1960}), \bibinfo{pages}{4--11}.
\newblock
\showISSN{0099-4561}
\urldef\tempurl%
\url{https://doi.org/10.1109/THFE2.1960.4503259}
\showDOI{\tempurl}


\bibitem[\protect\citeauthoryear{Lopes, Y{\"u}ksel, Guimbreti{\`e}re, and
  Baudisch}{Lopes et~al\mbox{.}}{2016}]%
        {lopesMuscleplotterInteractiveSystem2016}
\bibfield{author}{\bibinfo{person}{Pedro Lopes}, \bibinfo{person}{Do{\u a}a
  Y{\"u}ksel}, \bibinfo{person}{Fran{\c c}ois Guimbreti{\`e}re}, {and}
  \bibinfo{person}{Patrick Baudisch}.} \bibinfo{year}{2016}\natexlab{}.
\newblock \showarticletitle{Muscle-Plotter: An {{Interactive System}} Based on
  {{Electrical Muscle Stimulation}} That {{Produces Spatial Output}}}. In
  \bibinfo{booktitle}{\emph{Proceedings of the 29th {{Annual Symposium}} on
  {{User Interface Software}} and {{Technology}} - {{UIST}} '16}}.
  \bibinfo{publisher}{{ACM Press}}, \bibinfo{address}{{Tokyo, Japan}},
  \bibinfo{pages}{207--217}.
\newblock
\showISBNx{978-1-4503-4189-9}
\urldef\tempurl%
\url{https://doi.org/10.1145/2984511.2984530}
\showDOI{\tempurl}


\bibitem[\protect\citeauthoryear{Mahapatra, Jensen, McQuaid, and
  Ashbrook}{Mahapatra et~al\mbox{.}}{2019}]%
        {mahapatraBarriersEndUserDesigners2019}
\bibfield{author}{\bibinfo{person}{Chandan Mahapatra},
  \bibinfo{person}{Jonas~Kjeldmand Jensen}, \bibinfo{person}{Michael McQuaid},
  {and} \bibinfo{person}{Daniel Ashbrook}.} \bibinfo{year}{2019}\natexlab{}.
\newblock \showarticletitle{Barriers to {{End}}-{{User Designers}} of
  {{Augmented Fabrication}}}. In \bibinfo{booktitle}{\emph{Proceedings of the
  2019 {{CHI Conference}} on {{Human Factors}} in {{Computing Systems}}}}
  \emph{(\bibinfo{series}{{{CHI}} '19})}. \bibinfo{publisher}{{ACM}},
  \bibinfo{address}{{New York, NY, USA}}, \bibinfo{pages}{383:1--383:15}.
\newblock
\showISBNx{978-1-4503-5970-2}
\urldef\tempurl%
\url{https://doi.org/10.1145/3290605.3300613}
\showDOI{\tempurl}


\bibitem[\protect\citeauthoryear{Mateevitsi, Haggadone, Leigh, Kunzer, and
  Kenyon}{Mateevitsi et~al\mbox{.}}{2013}]%
        {mateevitsiSensingEnvironmentSpiderSense2013}
\bibfield{author}{\bibinfo{person}{Victor Mateevitsi}, \bibinfo{person}{Brad
  Haggadone}, \bibinfo{person}{Jason Leigh}, \bibinfo{person}{Brian Kunzer},
  {and} \bibinfo{person}{Robert~V. Kenyon}.} \bibinfo{year}{2013}\natexlab{}.
\newblock \showarticletitle{Sensing the {{Environment Through SpiderSense}}}.
  In \bibinfo{booktitle}{\emph{Proceedings of the 4th {{Augmented Human
  International Conference}}}} \emph{(\bibinfo{series}{{{AH}} '13})}.
  \bibinfo{publisher}{{ACM}}, \bibinfo{address}{{New York, NY, USA}},
  \bibinfo{pages}{51--57}.
\newblock
\showISBNx{978-1-4503-1904-1}
\urldef\tempurl%
\url{https://doi.org/10.1145/2459236.2459246}
\showDOI{\tempurl}


\bibitem[\protect\citeauthoryear{{Merriam-Webster
  Incorporated}}{{Merriam-Webster Incorporated}}{2019}]%
        {merriam-websterincorporatedDefinitionEphemeral2019}
\bibfield{author}{\bibinfo{person}{{Merriam-Webster Incorporated}}.}
  \bibinfo{year}{2019}\natexlab{}.
\newblock \bibinfo{title}{Definition of {{Ephemeral}}}.
\newblock
  \bibinfo{howpublished}{https://www.merriam-webster.com/dictionary/ephemeral}.
\newblock
\newblock
\shownote{(Accessed 27.07.2021)}.


\bibitem[\protect\citeauthoryear{Mitrovic}{Mitrovic}{2015}]%
        {mitrovicIntroductionSpeculativeDesign2015}
\bibfield{author}{\bibinfo{person}{Ivica Mitrovic}.}
  \bibinfo{year}{2015}\natexlab{}.
\newblock \bibinfo{booktitle}{\emph{Introduction to {{Speculative Design
  Practice}} \textendash{} {{Eutropia}}, a {{Case Study}}}}.
\newblock \bibinfo{publisher}{{HDD \& DVK UMAS}},
  \bibinfo{address}{{Zagreb/Split}}.
\newblock
\showISBNx{978-953-6778-10-2}


\bibitem[\protect\citeauthoryear{Mueller, Im, Gurevich, Teibrich, Pfisterer,
  Guimbreti{\`e}re, and Baudisch}{Mueller et~al\mbox{.}}{2014a}]%
        {muellerWirePrint3DPrinted2014}
\bibfield{author}{\bibinfo{person}{Stefanie Mueller}, \bibinfo{person}{Sangha
  Im}, \bibinfo{person}{Serafima Gurevich}, \bibinfo{person}{Alexander
  Teibrich}, \bibinfo{person}{Lisa Pfisterer}, \bibinfo{person}{Fran{\c c}ois
  Guimbreti{\`e}re}, {and} \bibinfo{person}{Patrick Baudisch}.}
  \bibinfo{year}{2014}\natexlab{a}.
\newblock \showarticletitle{{{WirePrint}}: {{3D Printed Previews}} for {{Fast
  Prototyping}}}. In \bibinfo{booktitle}{\emph{Proceedings of the 27th {{Annual
  ACM Symposium}} on {{User Interface Software}} and {{Technology}}}}
  \emph{(\bibinfo{series}{{{UIST}} '14})}. \bibinfo{publisher}{{ACM}},
  \bibinfo{address}{{New York, NY, USA}}, \bibinfo{pages}{273--280}.
\newblock
\showISBNx{978-1-4503-3069-5}
\urldef\tempurl%
\url{https://doi.org/10.1145/2642918.2647359}
\showDOI{\tempurl}


\bibitem[\protect\citeauthoryear{Mueller, Mohr, Guenther, Frohnhofen, and
  Baudisch}{Mueller et~al\mbox{.}}{2014b}]%
        {muellerFaBrickationFast3D2014}
\bibfield{author}{\bibinfo{person}{Stefanie Mueller}, \bibinfo{person}{Tobias
  Mohr}, \bibinfo{person}{Kerstin Guenther}, \bibinfo{person}{Johannes
  Frohnhofen}, {and} \bibinfo{person}{Patrick Baudisch}.}
  \bibinfo{year}{2014}\natexlab{b}.
\newblock \showarticletitle{{{faBrickation}}: Fast {{3D Printing}} of
  {{Functional Objects}} by {{Integrating Construction Kit Building Blocks}}}.
  In \bibinfo{booktitle}{\emph{Proceedings of the {{SIGCHI Conference}} on
  {{Human Factors}} in {{Computing Systems}}}} \emph{(\bibinfo{series}{{{CHI}}
  '14})}. \bibinfo{publisher}{{ACM}}, \bibinfo{address}{{New York, NY, USA}},
  \bibinfo{pages}{3827--3834}.
\newblock
\showISBNx{978-1-4503-2473-1}
\urldef\tempurl%
\url{https://doi.org/10.1145/2556288.2557005}
\showDOI{\tempurl}


\bibitem[\protect\citeauthoryear{Nabeshima, Saraiji, and Minamizawa}{Nabeshima
  et~al\mbox{.}}{2019}]%
        {nabeshimaProstheticTailArtificial2019}
\bibfield{author}{\bibinfo{person}{Junichi Nabeshima},
  \bibinfo{person}{MHD~Yamen Saraiji}, {and} \bibinfo{person}{Kouta
  Minamizawa}.} \bibinfo{year}{2019}\natexlab{}.
\newblock \showarticletitle{Prosthetic {{Tail}}: Artificial {{Anthropomorphic
  Tail}} for {{Extending Innate Body Functions}}}. In
  \bibinfo{booktitle}{\emph{Proceedings of the 10th {{Augmented Human
  International Conference}} 2019}} \emph{(\bibinfo{series}{{{AH2019}}})}.
  \bibinfo{publisher}{{ACM}}, \bibinfo{address}{{New York, NY, USA}},
  \bibinfo{pages}{36:1--36:4}.
\newblock
\showISBNx{978-1-4503-6547-5}
\urldef\tempurl%
\url{https://doi.org/10.1145/3311823.3311848}
\showDOI{\tempurl}


\bibitem[\protect\citeauthoryear{Napier}{Napier}{1956}]%
        {napierPrehensileMovementsHuman1956}
\bibfield{author}{\bibinfo{person}{J.~R. Napier}.}
  \bibinfo{year}{1956}\natexlab{}.
\newblock \showarticletitle{The Prehensile Movements of the Human Hand}.
\newblock \bibinfo{journal}{\emph{The Journal of Bone and Joint Surgery.
  British volume}} \bibinfo{volume}{38-B}, \bibinfo{number}{4}
  (\bibinfo{date}{Nov.} \bibinfo{year}{1956}), \bibinfo{pages}{902--913}.
\newblock
\showISSN{0301-620X}
\urldef\tempurl%
\url{https://doi.org/10.1302/0301-620X.38B4.902}
\showDOI{\tempurl}


\bibitem[\protect\citeauthoryear{Nisser, Zhu, Chen, Bulovic, Punpongsanon, and
  Mueller}{Nisser et~al\mbox{.}}{2019}]%
        {nisserSequentialSupport3D2019}
\bibfield{author}{\bibinfo{person}{Martin Nisser}, \bibinfo{person}{Junyi Zhu},
  \bibinfo{person}{Tianye Chen}, \bibinfo{person}{Katarina Bulovic},
  \bibinfo{person}{Parinya Punpongsanon}, {and} \bibinfo{person}{Stefanie
  Mueller}.} \bibinfo{year}{2019}\natexlab{}.
\newblock \showarticletitle{Sequential {{Support}}: {{3D Printing Dissolvable
  Support Material}} for {{Time}}-{{Dependent Mechanisms}}}. In
  \bibinfo{booktitle}{\emph{Proceedings of the {{Thirteenth International
  Conference}} on {{Tangible}}, {{Embedded}}, and {{Embodied Interaction}}}}
  \emph{(\bibinfo{series}{{{TEI}} '19})}. \bibinfo{publisher}{{ACM}},
  \bibinfo{address}{{New York, NY, USA}}, \bibinfo{pages}{669--676}.
\newblock
\showISBNx{978-1-4503-6196-5}
\urldef\tempurl%
\url{https://doi.org/10.1145/3294109.3295630}
\showDOI{\tempurl}


\bibitem[\protect\citeauthoryear{Orchard}{Orchard}{1998}]%
        {orchardCassellDictionaryNorse1998}
\bibfield{author}{\bibinfo{person}{Andy Orchard}.}
  \bibinfo{year}{1998}\natexlab{}.
\newblock \bibinfo{booktitle}{\emph{Cassell Dictionary of {{Norse}} Myth and
  Legend}}.
\newblock \bibinfo{publisher}{{Cassell}}, \bibinfo{address}{{Wellington House,
  125 Strand, London WC2R OBB}}.
\newblock
\showISBNx{0 304 34520 2}


\bibitem[\protect\citeauthoryear{Oxman, Laucks, Kayser, Tsai, and
  Firstenberg}{Oxman et~al\mbox{.}}{2013}]%
        {oxmanFreeform3DPrinting2013}
\bibfield{author}{\bibinfo{person}{Neri Oxman}, \bibinfo{person}{Jared Laucks},
  \bibinfo{person}{Markus Kayser}, \bibinfo{person}{Elizabeth Tsai}, {and}
  \bibinfo{person}{Michal Firstenberg}.} \bibinfo{year}{2013}\natexlab{}.
\newblock \showarticletitle{Freeform {{3D}} Printing: Towards a Sustainable
  Approach to Additive Manufacturing}.
\newblock \bibinfo{journal}{\emph{Green design, materials and manufacturing
  processes}}  \bibinfo{volume}{479} (\bibinfo{year}{2013}),
  \bibinfo{pages}{479--483}.
\newblock


\bibitem[\protect\citeauthoryear{Pal}{Pal}{2017}]%
        {palCHI4GoodGood4CHI2017}
\bibfield{author}{\bibinfo{person}{Joyojeet Pal}.}
  \bibinfo{year}{2017}\natexlab{}.
\newblock \showarticletitle{{{CHI4Good}} or {{Good4CHI}}}.
\newblock In \bibinfo{booktitle}{\emph{Proceedings of the 2017 {{CHI Conference
  Extended Abstracts}} on {{Human Factors}} in {{Computing Systems}}}}.
  \bibinfo{publisher}{{Association for Computing Machinery}},
  \bibinfo{address}{{New York, NY, USA}}, \bibinfo{pages}{709--721}.
\newblock
\showISBNx{978-1-4503-4656-6}


\bibitem[\protect\citeauthoryear{Pearce and Zalucki}{Pearce and
  Zalucki}{2002}]%
        {pearceSpiderBallooningCrops2002}
\bibfield{author}{\bibinfo{person}{Sarina Pearce} {and} \bibinfo{person}{M.~P.
  Zalucki}.} \bibinfo{year}{2002}\natexlab{}.
\newblock \showarticletitle{Spider Ballooning in Crops: A Web of Intrigue}.
\newblock \bibinfo{journal}{\emph{Australian Grain}} \bibinfo{volume}{12},
  \bibinfo{number}{4} (\bibinfo{year}{2002}), \bibinfo{pages}{vi--vii}.
\newblock


\bibitem[\protect\citeauthoryear{Peek and Moyer}{Peek and Moyer}{2017}]%
        {peekPopfabCasePortable2017}
\bibfield{author}{\bibinfo{person}{Nadya Peek} {and} \bibinfo{person}{Ilan
  Moyer}.} \bibinfo{year}{2017}\natexlab{}.
\newblock \showarticletitle{Popfab: A {{Case}} for {{Portable Digital
  Fabrication}}}. In \bibinfo{booktitle}{\emph{Proceedings of the {{Eleventh
  International Conference}} on {{Tangible}}, {{Embedded}}, and {{Embodied
  Interaction}}}} \emph{(\bibinfo{series}{{{TEI}} '17})}.
  \bibinfo{publisher}{{ACM}}, \bibinfo{address}{{New York, NY, USA}},
  \bibinfo{pages}{325--329}.
\newblock
\showISBNx{978-1-4503-4676-4}
\urldef\tempurl%
\url{https://doi.org/10.1145/3024969.3025009}
\showDOI{\tempurl}


\bibitem[\protect\citeauthoryear{Peng, Guimbreti{\`e}re, McCann, and
  Hudson}{Peng et~al\mbox{.}}{2016}]%
        {peng3DPrinterInteractive2016}
\bibfield{author}{\bibinfo{person}{Huaishu Peng}, \bibinfo{person}{Fran{\c
  c}ois Guimbreti{\`e}re}, \bibinfo{person}{James McCann}, {and}
  \bibinfo{person}{Scott Hudson}.} \bibinfo{year}{2016}\natexlab{}.
\newblock \showarticletitle{A {{3D Printer}} for {{Interactive Electromagnetic
  Devices}}}. In \bibinfo{booktitle}{\emph{Proceedings of the 29th {{Annual
  Symposium}} on {{User Interface Software}} and {{Technology}}}}
  \emph{(\bibinfo{series}{{{UIST}} '16})}. \bibinfo{publisher}{{Association for
  Computing Machinery}}, \bibinfo{address}{{New York, NY, USA}},
  \bibinfo{pages}{553--562}.
\newblock
\showISBNx{978-1-4503-4189-9}
\urldef\tempurl%
\url{https://doi.org/10.1145/2984511.2984523}
\showDOI{\tempurl}


\bibitem[\protect\citeauthoryear{Peng, Zoran, and Guimbreti{\`e}re}{Peng
  et~al\mbox{.}}{2015}]%
        {pengDCoilHandsonApproach2015}
\bibfield{author}{\bibinfo{person}{Huaishu Peng}, \bibinfo{person}{Amit Zoran},
  {and} \bibinfo{person}{Fran{\c c}ois~V. Guimbreti{\`e}re}.}
  \bibinfo{year}{2015}\natexlab{}.
\newblock \showarticletitle{D-{{Coil}}: A {{Hands}}-on {{Approach}} to
  {{Digital 3D Models Design}}}. In \bibinfo{booktitle}{\emph{Proceedings of
  the 33rd {{Annual ACM Conference}} on {{Human Factors}} in {{Computing
  Systems}}}} \emph{(\bibinfo{series}{{{CHI}} '15})}.
  \bibinfo{publisher}{{ACM}}, \bibinfo{address}{{New York, NY, USA}},
  \bibinfo{pages}{1807--1815}.
\newblock
\showISBNx{978-1-4503-3145-6}
\urldef\tempurl%
\url{https://doi.org/10.1145/2702123.2702381}
\showDOI{\tempurl}


\bibitem[\protect\citeauthoryear{Pusch, Hinton, and Feinberg}{Pusch
  et~al\mbox{.}}{2018}]%
        {puschLargeVolumeSyringe2018}
\bibfield{author}{\bibinfo{person}{Kira Pusch}, \bibinfo{person}{Thomas~J.
  Hinton}, {and} \bibinfo{person}{Adam~W. Feinberg}.}
  \bibinfo{year}{2018}\natexlab{}.
\newblock \showarticletitle{Large Volume Syringe Pump Extruder for Desktop
  {{3D}} Printers}.
\newblock \bibinfo{journal}{\emph{HardwareX}}  \bibinfo{volume}{3}
  (\bibinfo{date}{April} \bibinfo{year}{2018}), \bibinfo{pages}{49--61}.
\newblock
\showISSN{2468-0672}
\urldef\tempurl%
\url{https://doi.org/10.1016/j.ohx.2018.02.001}
\showDOI{\tempurl}


\bibitem[\protect\citeauthoryear{Quitmeyer and {Perner-Wilson}}{Quitmeyer and
  {Perner-Wilson}}{2015}]%
        {quitmeyerWearableStudioPractice2015}
\bibfield{author}{\bibinfo{person}{Andrew Quitmeyer} {and}
  \bibinfo{person}{Hannah {Perner-Wilson}}.} \bibinfo{year}{2015}\natexlab{}.
\newblock \showarticletitle{Wearable {{Studio Practice}}: Design
  {{Considerations}} for {{Digital Crafting}} in {{Harsh Environments}}}. In
  \bibinfo{booktitle}{\emph{Adjunct {{Proceedings}} of the 2015 {{ACM
  International Joint Conference}} on {{Pervasive}} and {{Ubiquitous
  Computing}} and {{Proceedings}} of the 2015 {{ACM International Symposium}}
  on {{Wearable Computers}}}} \emph{(\bibinfo{series}{{{UbiComp}}/{{ISWC}}'15
  {{Adjunct}}})}. \bibinfo{publisher}{{ACM}}, \bibinfo{address}{{New York, NY,
  USA}}, \bibinfo{pages}{1285--1293}.
\newblock
\showISBNx{978-1-4503-3575-1}
\urldef\tempurl%
\url{https://doi.org/10.1145/2800835.2807926}
\showDOI{\tempurl}


\bibitem[\protect\citeauthoryear{Rhodes and Starner}{Rhodes and
  Starner}{1996}]%
        {rhodesRemembranceAgentContinuously1996}
\bibfield{author}{\bibinfo{person}{Bradley~J Rhodes} {and}
  \bibinfo{person}{Thad Starner}.} \bibinfo{year}{1996}\natexlab{}.
\newblock \showarticletitle{Remembrance {{Agent}}: A {{Continuously Running
  Automated Information Retrieval System}}}.
\newblock \bibinfo{journal}{\emph{The Proceedings of The First International
  Conference on The Practical Application Of Intelligent Agents and Multi Agent
  Technology}} \bibinfo{volume}{1}, \bibinfo{number}{1} (\bibinfo{year}{1996}),
  \bibinfo{pages}{487--495}.
\newblock


\bibitem[\protect\citeauthoryear{Roumen, Kruck, D{\"u}rschmid, Nack, and
  Baudisch}{Roumen et~al\mbox{.}}{2016}]%
        {roumenMobileFabrication2016}
\bibfield{author}{\bibinfo{person}{Thijs Roumen}, \bibinfo{person}{Bastian
  Kruck}, \bibinfo{person}{Tobias D{\"u}rschmid}, \bibinfo{person}{Tobias
  Nack}, {and} \bibinfo{person}{Patrick Baudisch}.}
  \bibinfo{year}{2016}\natexlab{}.
\newblock \showarticletitle{Mobile {{Fabrication}}}. In
  \bibinfo{booktitle}{\emph{Proceedings of the 29th {{Annual Symposium}} on
  {{User Interface Software}} and {{Technology}}}}
  \emph{(\bibinfo{series}{{{UIST}} '16})}. \bibinfo{publisher}{{ACM}},
  \bibinfo{address}{{New York, NY, USA}}, \bibinfo{pages}{3--14}.
\newblock
\showISBNx{978-1-4503-4189-9}
\urldef\tempurl%
\url{https://doi.org/10.1145/2984511.2984586}
\showDOI{\tempurl}


\bibitem[\protect\citeauthoryear{Saraiji, Sasaki, Kunze, Minamizawa, and
  Inami}{Saraiji et~al\mbox{.}}{2018}]%
        {saraijiMetaArmsBodyRemapping2018}
\bibfield{author}{\bibinfo{person}{MHD~Yamen Saraiji}, \bibinfo{person}{Tomoya
  Sasaki}, \bibinfo{person}{Kai Kunze}, \bibinfo{person}{Kouta Minamizawa},
  {and} \bibinfo{person}{Masahiko Inami}.} \bibinfo{year}{2018}\natexlab{}.
\newblock \showarticletitle{{{MetaArms}}: Body {{Remapping Using
  Feet}}-{{Controlled Artificial Arms}}}. In
  \bibinfo{booktitle}{\emph{Proceedings of the 31st {{Annual ACM Symposium}} on
  {{User Interface Software}} and {{Technology}}}}
  \emph{(\bibinfo{series}{{{UIST}} '18})}. \bibinfo{publisher}{{ACM}},
  \bibinfo{address}{{New York, NY, USA}}, \bibinfo{pages}{65--74}.
\newblock
\showISBNx{978-1-4503-5948-1}
\urldef\tempurl%
\url{https://doi.org/10.1145/3242587.3242665}
\showDOI{\tempurl}


\bibitem[\protect\citeauthoryear{Sasaki, Saraiji, Fernando, Minamizawa, and
  Inami}{Sasaki et~al\mbox{.}}{2017}]%
        {sasakiMetaLimbsMultipleArms2017}
\bibfield{author}{\bibinfo{person}{Tomoya Sasaki}, \bibinfo{person}{MHD~Yamen
  Saraiji}, \bibinfo{person}{Charith~Lasantha Fernando}, \bibinfo{person}{Kouta
  Minamizawa}, {and} \bibinfo{person}{Masahiko Inami}.}
  \bibinfo{year}{2017}\natexlab{}.
\newblock \showarticletitle{{{MetaLimbs}}: Multiple {{Arms Interaction
  Metamorphism}}}. In \bibinfo{booktitle}{\emph{{{ACM SIGGRAPH}} 2017
  {{Emerging Technologies}}}} \emph{(\bibinfo{series}{{{SIGGRAPH}} '17})}.
  \bibinfo{publisher}{{ACM}}, \bibinfo{address}{{New York, NY, USA}},
  \bibinfo{pages}{16:1--16:2}.
\newblock
\showISBNx{978-1-4503-5012-9}
\urldef\tempurl%
\url{https://doi.org/10.1145/3084822.3084837}
\showDOI{\tempurl}


\bibitem[\protect\citeauthoryear{Savage, Follmer, Li, and Hartmann}{Savage
  et~al\mbox{.}}{2015}]%
        {savageMakersMarksPhysical2015}
\bibfield{author}{\bibinfo{person}{Valkyrie Savage}, \bibinfo{person}{Sean
  Follmer}, \bibinfo{person}{Jingyi Li}, {and} \bibinfo{person}{Bj{\"o}rn
  Hartmann}.} \bibinfo{year}{2015}\natexlab{}.
\newblock \showarticletitle{Makers' {{Marks}}: Physical {{Markup}} for
  {{Designing}} and {{Fabricating Functional Objects}}}. In
  \bibinfo{booktitle}{\emph{Proceedings of the 28th {{Annual ACM Symposium}} on
  {{User Interface Software}} \& {{Technology}}}}
  \emph{(\bibinfo{series}{{{UIST}} '15})}. \bibinfo{publisher}{{ACM}},
  \bibinfo{address}{{New York, NY, USA}}, \bibinfo{pages}{103--108}.
\newblock
\showISBNx{978-1-4503-3779-3}
\urldef\tempurl%
\url{https://doi.org/10.1145/2807442.2807508}
\showDOI{\tempurl}


\bibitem[\protect\citeauthoryear{Schmidt}{Schmidt}{2017}]%
        {schmidtTechnologiesAmplifyMind2017}
\bibfield{author}{\bibinfo{person}{Albrecht Schmidt}.}
  \bibinfo{year}{2017}\natexlab{}.
\newblock \showarticletitle{Technologies to {{Amplify}} the {{Mind}}}.
\newblock \bibinfo{journal}{\emph{Computer}} \bibinfo{volume}{50},
  \bibinfo{number}{10} (\bibinfo{year}{2017}), \bibinfo{pages}{102--106}.
\newblock
\showISSN{0018-9162}
\urldef\tempurl%
\url{https://doi.org/10.1109/MC.2017.3641644}
\showDOI{\tempurl}


\bibitem[\protect\citeauthoryear{Shen, Dou, and Gu}{Shen et~al\mbox{.}}{2013}]%
        {shenRoCuModelIterativeTangible2013}
\bibfield{author}{\bibinfo{person}{Yuebo Shen}, \bibinfo{person}{Keqin Dou},
  {and} \bibinfo{person}{Jiawei Gu}.} \bibinfo{year}{2013}\natexlab{}.
\newblock \showarticletitle{{{RoCuModel}}: An {{Iterative Tangible Modeling
  System}}}. In \bibinfo{booktitle}{\emph{Proceedings of the 8th
  {{International Conference}} on {{Tangible}}, {{Embedded}} and {{Embodied
  Interaction}}}} \emph{(\bibinfo{series}{{{TEI}} '14})}.
  \bibinfo{publisher}{{ACM}}, \bibinfo{address}{{New York, NY, USA}},
  \bibinfo{pages}{73--76}.
\newblock
\showISBNx{978-1-4503-2635-3}
\urldef\tempurl%
\url{https://doi.org/10.1145/2540930.2540960}
\showDOI{\tempurl}


\bibitem[\protect\citeauthoryear{Shilkrot, Huber, Meng~Ee, Maes, and
  Nanayakkara}{Shilkrot et~al\mbox{.}}{2015a}]%
        {shilkrotFingerReaderWearableDevice2015}
\bibfield{author}{\bibinfo{person}{Roy Shilkrot}, \bibinfo{person}{Jochen
  Huber}, \bibinfo{person}{Wong Meng~Ee}, \bibinfo{person}{Pattie Maes}, {and}
  \bibinfo{person}{Suranga~Chandima Nanayakkara}.}
  \bibinfo{year}{2015}\natexlab{a}.
\newblock \showarticletitle{{{FingerReader}}: A {{Wearable Device}} to
  {{Explore Printed Text}} on the {{Go}}}. In
  \bibinfo{booktitle}{\emph{Proceedings of the 33rd {{Annual ACM Conference}}
  on {{Human Factors}} in {{Computing Systems}}}}
  \emph{(\bibinfo{series}{{{CHI}} '15})}. \bibinfo{publisher}{{ACM}},
  \bibinfo{address}{{New York, NY, USA}}, \bibinfo{pages}{2363--2372}.
\newblock
\showISBNx{978-1-4503-3145-6}
\urldef\tempurl%
\url{https://doi.org/10.1145/2702123.2702421}
\showDOI{\tempurl}


\bibitem[\protect\citeauthoryear{Shilkrot, Huber, Steimle, Nanayakkara, and
  Maes}{Shilkrot et~al\mbox{.}}{2015b}]%
        {shilkrotDigitalDigitsComprehensive2015}
\bibfield{author}{\bibinfo{person}{Roy Shilkrot}, \bibinfo{person}{Jochen
  Huber}, \bibinfo{person}{J{\"u}rgen Steimle}, \bibinfo{person}{Suranga
  Nanayakkara}, {and} \bibinfo{person}{Pattie Maes}.}
  \bibinfo{year}{2015}\natexlab{b}.
\newblock \showarticletitle{Digital {{Digits}}: A {{Comprehensive Survey}} of
  {{Finger Augmentation Devices}}}.
\newblock \bibinfo{journal}{\emph{ACM Comput. Surv.}} \bibinfo{volume}{48},
  \bibinfo{number}{2} (\bibinfo{date}{Nov.} \bibinfo{year}{2015}),
  \bibinfo{pages}{30:1--30:29}.
\newblock
\showISSN{0360-0300}
\urldef\tempurl%
\url{https://doi.org/10.1145/2828993}
\showDOI{\tempurl}


\bibitem[\protect\citeauthoryear{Shneiderman}{Shneiderman}{1999}]%
        {shneidermanHumanValuesFuture1999}
\bibfield{author}{\bibinfo{person}{Ben Shneiderman}.}
  \bibinfo{year}{1999}\natexlab{}.
\newblock \showarticletitle{Human Values and the Future of Technology: A
  Declaration of Responsibility}.
\newblock \bibinfo{journal}{\emph{ACM SIGCAS Computers and Society}}
  \bibinfo{volume}{29}, \bibinfo{number}{3} (\bibinfo{year}{1999}),
  \bibinfo{pages}{5--9}.
\newblock
\urldef\tempurl%
\url{https://doi.org/10.1145/572183.572185}
\showDOI{\tempurl}


\bibitem[\protect\citeauthoryear{Song and Paulos}{Song and Paulos}{2021}]%
        {songUnmakingEnablingCelebrating2021}
\bibfield{author}{\bibinfo{person}{Katherine~W Song} {and}
  \bibinfo{person}{Eric Paulos}.} \bibinfo{year}{2021}\natexlab{}.
\newblock \showarticletitle{Unmaking: Enabling and {{Celebrating}} the
  {{Creative Material}} of {{Failure}}, {{Destruction}}, {{Decay}}, and
  {{Deformation}}}. In \bibinfo{booktitle}{\emph{Proceedings of the 2021 {{CHI
  Conference}} on {{Human Factors}} in {{Computing Systems}}}}
  \emph{(\bibinfo{series}{{{CHI}} '21})}. \bibinfo{publisher}{{Association for
  Computing Machinery}}, \bibinfo{address}{{New York, NY, USA}},
  \bibinfo{pages}{1--12}.
\newblock
\showISBNx{978-1-4503-8096-6}
\urldef\tempurl%
\url{https://doi.org/10.1145/3411764.3445529}
\showDOI{\tempurl}


\bibitem[\protect\citeauthoryear{Stemasov}{Stemasov}{2021}]%
        {stemasovEnablingUbiquitousPersonal2021}
\bibfield{author}{\bibinfo{person}{Evgeny Stemasov}.}
  \bibinfo{year}{2021}\natexlab{}.
\newblock \showarticletitle{Enabling {{Ubiquitous Personal Fabrication}} by
  {{Deconstructing Established Notions}} of {{Artifact Modeling}}}. In
  \bibinfo{booktitle}{\emph{The {{Adjunct Publication}} of the 34th {{Annual
  ACM Symposium}} on {{User Interface Software}} and {{Technology}}}}
  \emph{(\bibinfo{series}{{{UIST}} '21})}. \bibinfo{publisher}{{Association for
  Computing Machinery}}, \bibinfo{address}{{New York, NY, USA}},
  \bibinfo{pages}{166--170}.
\newblock
\showISBNx{978-1-4503-8655-5}
\urldef\tempurl%
\url{https://doi.org/10.1145/3474349.3477589}
\showDOI{\tempurl}


\bibitem[\protect\citeauthoryear{Stemasov, Rukzio, and Gugenheimer}{Stemasov
  et~al\mbox{.}}{2021}]%
        {stemasovRoadUbiquitousPersonal2021}
\bibfield{author}{\bibinfo{person}{Evgeny Stemasov}, \bibinfo{person}{Enrico
  Rukzio}, {and} \bibinfo{person}{Jan Gugenheimer}.}
  \bibinfo{year}{2021}\natexlab{}.
\newblock \showarticletitle{The {{Road}} to {{Ubiquitous Personal
  Fabrication}}: Modeling-{{Free Instead}} of {{Increasingly Simple}}}.
\newblock \bibinfo{journal}{\emph{IEEE Pervasive Computing}}
  \bibinfo{volume}{20}, \bibinfo{number}{1} (\bibinfo{year}{2021}),
  \bibinfo{pages}{1--9}.
\newblock
\showISSN{1558-2590}
\urldef\tempurl%
\url{https://doi.org/10.1109/MPRV.2020.3029650}
\showDOI{\tempurl}


\bibitem[\protect\citeauthoryear{Stemasov, Wagner, Gugenheimer, and
  Rukzio}{Stemasov et~al\mbox{.}}{2020}]%
        {stemasovMixMatchOmitting2020}
\bibfield{author}{\bibinfo{person}{Evgeny Stemasov}, \bibinfo{person}{Tobias
  Wagner}, \bibinfo{person}{Jan Gugenheimer}, {and} \bibinfo{person}{Enrico
  Rukzio}.} \bibinfo{year}{2020}\natexlab{}.
\newblock \showarticletitle{Mix\&{{Match}}: Towards {{Omitting Modelling
  Through In}}-Situ {{Remixing}} of {{Model Repository Artifacts}} in {{Mixed
  Reality}}}. In \bibinfo{booktitle}{\emph{Proceedings of the 2020 {{CHI
  Conference}} on {{Human Factors}} in {{Computing Systems}}}}
  \emph{(\bibinfo{series}{{{CHI}} '20})}. \bibinfo{publisher}{{Association for
  Computing Machinery}}, \bibinfo{address}{{Honolulu, HI, USA}},
  \bibinfo{pages}{1--12}.
\newblock
\showISBNx{978-1-4503-6708-0}
\urldef\tempurl%
\url{https://doi.org/10.1145/3313831.3376839}
\showDOI{\tempurl}


\bibitem[\protect\citeauthoryear{Tanenbaum}{Tanenbaum}{2014}]%
        {tanenbaumDesignFictionalInteractions2014}
\bibfield{author}{\bibinfo{person}{Theresa~Jean Tanenbaum}.}
  \bibinfo{year}{2014}\natexlab{}.
\newblock \showarticletitle{Design Fictional Interactions: Why {{HCI}} Should
  Care about Stories}.
\newblock \bibinfo{journal}{\emph{Interactions}} \bibinfo{volume}{21},
  \bibinfo{number}{5} (\bibinfo{date}{Sept.} \bibinfo{year}{2014}),
  \bibinfo{pages}{22--23}.
\newblock
\showISSN{1072-5520}
\urldef\tempurl%
\url{https://doi.org/10.1145/2648414}
\showDOI{\tempurl}


\bibitem[\protect\citeauthoryear{Tao, Wang, Zhang, Lu, Zhang, Yao, and
  Ying}{Tao et~al\mbox{.}}{2017}]%
        {taoWeaveMeshLowFidelityLowCost2017}
\bibfield{author}{\bibinfo{person}{Ye Tao}, \bibinfo{person}{Guanyun Wang},
  \bibinfo{person}{Caowei Zhang}, \bibinfo{person}{Nannan Lu},
  \bibinfo{person}{Xiaolian Zhang}, \bibinfo{person}{Cheng Yao}, {and}
  \bibinfo{person}{Fangtian Ying}.} \bibinfo{year}{2017}\natexlab{}.
\newblock \showarticletitle{{{WeaveMesh}}: A {{Low}}-{{Fidelity}} and
  {{Low}}-{{Cost Prototyping Approach}} for {{3D Models Created}} by {{Flexible
  Assembly}}}. In \bibinfo{booktitle}{\emph{Proceedings of the 2017 {{CHI
  Conference}} on {{Human Factors}} in {{Computing Systems}}}}
  \emph{(\bibinfo{series}{{{CHI}} '17})}. \bibinfo{publisher}{{ACM}},
  \bibinfo{address}{{New York, NY, USA}}, \bibinfo{pages}{509--518}.
\newblock
\showISBNx{978-1-4503-4655-9}
\urldef\tempurl%
\url{https://doi.org/10.1145/3025453.3025699}
\showDOI{\tempurl}


\bibitem[\protect\citeauthoryear{Teibrich, Mueller, Guimbreti{\`e}re, Kovacs,
  Neubert, and Baudisch}{Teibrich et~al\mbox{.}}{2015}]%
        {teibrichPatchingPhysicalObjects2015}
\bibfield{author}{\bibinfo{person}{Alexander Teibrich},
  \bibinfo{person}{Stefanie Mueller}, \bibinfo{person}{Fran{\c c}ois
  Guimbreti{\`e}re}, \bibinfo{person}{Robert Kovacs}, \bibinfo{person}{Stefan
  Neubert}, {and} \bibinfo{person}{Patrick Baudisch}.}
  \bibinfo{year}{2015}\natexlab{}.
\newblock \showarticletitle{Patching {{Physical Objects}}}. In
  \bibinfo{booktitle}{\emph{Proceedings of the 28th {{Annual ACM Symposium}} on
  {{User Interface Software}} \& {{Technology}}}}
  \emph{(\bibinfo{series}{{{UIST}} '15})}. \bibinfo{publisher}{{ACM}},
  \bibinfo{address}{{New York, NY, USA}}, \bibinfo{pages}{83--91}.
\newblock
\showISBNx{978-1-4503-3779-3}
\urldef\tempurl%
\url{https://doi.org/10.1145/2807442.2807467}
\showDOI{\tempurl}


\bibitem[\protect\citeauthoryear{Vasquez and Vega}{Vasquez and Vega}{2019a}]%
        {vasquezPlasticBiomaterialsPrototyping2019}
\bibfield{author}{\bibinfo{person}{Eldy S.~Lazaro Vasquez} {and}
  \bibinfo{person}{Katia Vega}.} \bibinfo{year}{2019}\natexlab{a}.
\newblock \showarticletitle{From Plastic to Biomaterials: Prototyping {{DIY}}
  Electronics with Mycelium}. In \bibinfo{booktitle}{\emph{Adjunct
  {{Proceedings}} of the 2019 {{ACM International Joint Conference}} on
  {{Pervasive}} and {{Ubiquitous Computing}} and {{Proceedings}} of the 2019
  {{ACM International Symposium}} on {{Wearable Computers}}}}.
  \bibinfo{publisher}{{ACM}}, \bibinfo{address}{{London United Kingdom}},
  \bibinfo{pages}{308--311}.
\newblock
\showISBNx{978-1-4503-6869-8}
\urldef\tempurl%
\url{https://doi.org/10.1145/3341162.3343808}
\showDOI{\tempurl}


\bibitem[\protect\citeauthoryear{Vasquez and Vega}{Vasquez and Vega}{2019b}]%
        {vasquezMycoaccessoriesSustainableWearables2019}
\bibfield{author}{\bibinfo{person}{Eldy S.~Lazaro Vasquez} {and}
  \bibinfo{person}{Katia Vega}.} \bibinfo{year}{2019}\natexlab{b}.
\newblock \showarticletitle{Myco-Accessories: Sustainable Wearables with
  Biodegradable Materials}. In \bibinfo{booktitle}{\emph{Proceedings of the
  23rd {{International Symposium}} on {{Wearable Computers}}}}.
  \bibinfo{publisher}{{ACM}}, \bibinfo{address}{{London United Kingdom}},
  \bibinfo{pages}{306--311}.
\newblock
\showISBNx{978-1-4503-6870-4}
\urldef\tempurl%
\url{https://doi.org/10.1145/3341163.3346938}
\showDOI{\tempurl}


\bibitem[\protect\citeauthoryear{Vasquez, Wang, and Vega}{Vasquez
  et~al\mbox{.}}{2020}]%
        {vasquezEnvironmentalImpactPhysical2020}
\bibfield{author}{\bibinfo{person}{Eldy S~Lazaro Vasquez},
  \bibinfo{person}{Hao-Chuan Wang}, {and} \bibinfo{person}{Katia Vega}.}
  \bibinfo{year}{2020}\natexlab{}.
\newblock \showarticletitle{The {{Environmental Impact}} of {{Physical
  Prototyping}}: A {{Five}}-{{Year CHI Review}}}.
\newblock \bibinfo{journal}{\emph{Self-Sustainable CHI Workshop}}
  \bibinfo{volume}{1}, \bibinfo{number}{1} (\bibinfo{year}{2020}),
  \bibinfo{pages}{8}.
\newblock


\bibitem[\protect\citeauthoryear{Vogel, Campreguer~Fran{\c c}a, Economidou,
  Maurer, and Tscheligi}{Vogel et~al\mbox{.}}{2020}]%
        {vogelCircularHCITools2020}
\bibfield{author}{\bibinfo{person}{Susanna Vogel}, \bibinfo{person}{Nathalia
  Campreguer~Fran{\c c}a}, \bibinfo{person}{Eleni Economidou},
  \bibinfo{person}{Bernhard Maurer}, {and} \bibinfo{person}{Manfred
  Tscheligi}.} \bibinfo{year}{2020}\natexlab{}.
\newblock \showarticletitle{Circular {{HCI}}: Tools for {{Embedding Circular
  Thinking}} in {{Material}}-{{Driven Design}}}. In
  \bibinfo{booktitle}{\emph{Companion {{Publication}} of the 2020 {{ACM
  Designing Interactive Systems Conference}}}} \emph{(\bibinfo{series}{{{DIS}}'
  20 {{Companion}}})}. \bibinfo{publisher}{{Association for Computing
  Machinery}}, \bibinfo{address}{{New York, NY, USA}},
  \bibinfo{pages}{233--237}.
\newblock
\showISBNx{978-1-4503-7987-8}
\urldef\tempurl%
\url{https://doi.org/10.1145/3393914.3395894}
\showDOI{\tempurl}


\bibitem[\protect\citeauthoryear{Von~Frisch and Von~Frisch}{Von~Frisch and
  Von~Frisch}{1974}]%
        {vonfrischAnimalArchitecture1974}
\bibfield{author}{\bibinfo{person}{Karl Von~Frisch} {and} \bibinfo{person}{Otto
  Von~Frisch}.} \bibinfo{year}{1974}\natexlab{}.
\newblock \bibinfo{booktitle}{\emph{Animal Architecture}}.
\newblock \bibinfo{publisher}{{Harcourt Brace Jovanovich New York}},
  \bibinfo{address}{{London, UK}}.
\newblock


\bibitem[\protect\citeauthoryear{Wall, Jacobson, Vogel, and Schneider}{Wall
  et~al\mbox{.}}{2021}]%
        {wallScrappyUsingScrap2021}
\bibfield{author}{\bibinfo{person}{Ludwig~Wilhelm Wall}, \bibinfo{person}{Alec
  Jacobson}, \bibinfo{person}{Daniel Vogel}, {and} \bibinfo{person}{Oliver
  Schneider}.} \bibinfo{year}{2021}\natexlab{}.
\newblock \showarticletitle{Scrappy: Using {{Scrap Material}} as {{Infill}} to
  {{Make Fabrication More Sustainable}}}. In
  \bibinfo{booktitle}{\emph{Proceedings of the 2021 {{CHI Conference}} on
  {{Human Factors}} in {{Computing Systems}}}}.
  \bibinfo{publisher}{{Association for Computing Machinery}},
  \bibinfo{address}{{New York, NY, USA}}, \bibinfo{pages}{1--12}.
\newblock
\showISBNx{978-1-4503-8096-6}


\bibitem[\protect\citeauthoryear{Weichel, Hardy, Alexander, and
  Gellersen}{Weichel et~al\mbox{.}}{2015}]%
        {weichelReFormIntegratingPhysical2015}
\bibfield{author}{\bibinfo{person}{Christian Weichel}, \bibinfo{person}{John
  Hardy}, \bibinfo{person}{Jason Alexander}, {and} \bibinfo{person}{Hans
  Gellersen}.} \bibinfo{year}{2015}\natexlab{}.
\newblock \showarticletitle{{{ReForm}}: Integrating {{Physical}} and {{Digital
  Design Through Bidirectional Fabrication}}}. In
  \bibinfo{booktitle}{\emph{Proceedings of the 28th {{Annual ACM Symposium}} on
  {{User Interface Software}} \& {{Technology}}}}
  \emph{(\bibinfo{series}{{{UIST}} '15})}. \bibinfo{publisher}{{ACM}},
  \bibinfo{address}{{New York, NY, USA}}, \bibinfo{pages}{93--102}.
\newblock
\showISBNx{978-1-4503-3779-3}
\urldef\tempurl%
\url{https://doi.org/10.1145/2807442.2807451}
\showDOI{\tempurl}


\bibitem[\protect\citeauthoryear{Weiser}{Weiser}{1991}]%
        {weiserComputer21St1991}
\bibfield{author}{\bibinfo{person}{Mark Weiser}.}
  \bibinfo{year}{1991}\natexlab{}.
\newblock \showarticletitle{The {{Computer}} for the 21 St {{Century}}}.
\newblock \bibinfo{journal}{\emph{Scientific american}} \bibinfo{volume}{265},
  \bibinfo{number}{3} (\bibinfo{year}{1991}), \bibinfo{pages}{94--105}.
\newblock


\bibitem[\protect\citeauthoryear{Weiser and Brown}{Weiser and Brown}{1997}]%
        {weiserComingAgeCalm1997}
\bibfield{author}{\bibinfo{person}{Mark Weiser} {and}
  \bibinfo{person}{John~Seely Brown}.} \bibinfo{year}{1997}\natexlab{}.
\newblock \showarticletitle{The {{Coming Age}} of {{Calm Technology}}}.
\newblock In \bibinfo{booktitle}{\emph{Beyond {{Calculation}}}}.
  \bibinfo{publisher}{{Springer}}, \bibinfo{address}{{New York, NY, USA}},
  \bibinfo{pages}{75--85}.
\newblock


\bibitem[\protect\citeauthoryear{Wu and Devendorf}{Wu and Devendorf}{2020}]%
        {wuUnfabricateDesigningSmart2020}
\bibfield{author}{\bibinfo{person}{Shanel Wu} {and} \bibinfo{person}{Laura
  Devendorf}.} \bibinfo{year}{2020}\natexlab{}.
\newblock \showarticletitle{Unfabricate: Designing {{Smart Textiles}} for
  {{Disassembly}}}. In \bibinfo{booktitle}{\emph{Proceedings of the 2020 {{CHI
  Conference}} on {{Human Factors}} in {{Computing Systems}}}}
  \emph{(\bibinfo{series}{{{CHI}} '20})}. \bibinfo{publisher}{{Association for
  Computing Machinery}}, \bibinfo{address}{{New York, NY, USA}},
  \bibinfo{pages}{1--14}.
\newblock
\showISBNx{978-1-4503-6708-0}
\urldef\tempurl%
\url{https://doi.org/10.1145/3313831.3376227}
\showDOI{\tempurl}


\bibitem[\protect\citeauthoryear{Yamada, Morishige, Nozaki, Ogawa, Yonezawa,
  and Tokuda}{Yamada et~al\mbox{.}}{2016}]%
        {yamadaReFabricatorIntegratingEveryday2016}
\bibfield{author}{\bibinfo{person}{Suguru Yamada}, \bibinfo{person}{Hironao
  Morishige}, \bibinfo{person}{Hiroki Nozaki}, \bibinfo{person}{Masaki Ogawa},
  \bibinfo{person}{Takuro Yonezawa}, {and} \bibinfo{person}{Hideyuki Tokuda}.}
  \bibinfo{year}{2016}\natexlab{}.
\newblock \showarticletitle{{{ReFabricator}}: Integrating {{Everyday Objects}}
  for {{Digital Fabrication}}}. In \bibinfo{booktitle}{\emph{Proceedings of the
  2016 {{CHI Conference Extended Abstracts}} on {{Human Factors}} in
  {{Computing Systems}}}} \emph{(\bibinfo{series}{{{CHI EA}} '16})}.
  \bibinfo{publisher}{{ACM}}, \bibinfo{address}{{New York, NY, USA}},
  \bibinfo{pages}{3804--3807}.
\newblock
\showISBNx{978-1-4503-4082-3}
\urldef\tempurl%
\url{https://doi.org/10.1145/2851581.2890237}
\showDOI{\tempurl}


\bibitem[\protect\citeauthoryear{Yamaoka, Niiyama, and Kakehi}{Yamaoka
  et~al\mbox{.}}{2017}]%
        {yamaokaBlowFabRapidPrototyping2017}
\bibfield{author}{\bibinfo{person}{Junichi Yamaoka}, \bibinfo{person}{Ryuma
  Niiyama}, {and} \bibinfo{person}{Yasuaki Kakehi}.}
  \bibinfo{year}{2017}\natexlab{}.
\newblock \showarticletitle{{{BlowFab}}: Rapid {{Prototyping}} for {{Rigid}}
  and {{Reusable Objects Using Inflation}} of {{Laser}}-Cut {{Surfaces}}}. In
  \bibinfo{booktitle}{\emph{Proceedings of the 30th {{Annual ACM Symposium}} on
  {{User Interface Software}} and {{Technology}}}}
  \emph{(\bibinfo{series}{{{UIST}} '17})}. \bibinfo{publisher}{{ACM}},
  \bibinfo{address}{{New York, NY, USA}}, \bibinfo{pages}{461--469}.
\newblock
\showISBNx{978-1-4503-4981-9}
\urldef\tempurl%
\url{https://doi.org/10.1145/3126594.3126624}
\showDOI{\tempurl}


\bibitem[\protect\citeauthoryear{Yasu}{Yasu}{2016}]%
        {yasuMOR4RHowCreate2016}
\bibfield{author}{\bibinfo{person}{Kentaro Yasu}.}
  \bibinfo{year}{2016}\natexlab{}.
\newblock \showarticletitle{{{MOR4R}}: How to {{Create 3D Objects Using}} a
  {{Microwave Oven}}}. In \bibinfo{booktitle}{\emph{Proceedings of the {{TEI}}
  '16: Tenth {{International Conference}} on {{Tangible}}, {{Embedded}}, and
  {{Embodied Interaction}}}} \emph{(\bibinfo{series}{{{TEI}} '16})}.
  \bibinfo{publisher}{{ACM}}, \bibinfo{address}{{New York, NY, USA}},
  \bibinfo{pages}{365--372}.
\newblock
\showISBNx{978-1-4503-3582-9}
\urldef\tempurl%
\url{https://doi.org/10.1145/2839462.2839507}
\showDOI{\tempurl}


\bibitem[\protect\citeauthoryear{Yung, Li, and Ashbrook}{Yung
  et~al\mbox{.}}{2018}]%
        {yungPrinty3DInsituTangible2018}
\bibfield{author}{\bibinfo{person}{Amanda~K. Yung}, \bibinfo{person}{Zhiyuan
  Li}, {and} \bibinfo{person}{Daniel Ashbrook}.}
  \bibinfo{year}{2018}\natexlab{}.
\newblock \showarticletitle{{{Printy3D}}: In-Situ {{Tangible
  Three}}-Dimensional {{Design}} for {{Augmented Fabrication}}}. In
  \bibinfo{booktitle}{\emph{Proceedings of the 17th {{ACM Conference}} on
  {{Interaction Design}} and {{Children}}}} \emph{(\bibinfo{series}{{{IDC}}
  '18})}. \bibinfo{publisher}{{ACM}}, \bibinfo{address}{{New York, NY, USA}},
  \bibinfo{pages}{181--194}.
\newblock
\showISBNx{978-1-4503-5152-2}
\urldef\tempurl%
\url{https://doi.org/10.1145/3202185.3202751}
\showDOI{\tempurl}


\bibitem[\protect\citeauthoryear{Zhu, Li, Mohideen, Hu, Gupta, Ramakrishna, and
  Liu}{Zhu et~al\mbox{.}}{2021}]%
        {zhuRealizationCircularEconomy2021}
\bibfield{author}{\bibinfo{person}{Caihan Zhu}, \bibinfo{person}{Tianya Li},
  \bibinfo{person}{Mohamedazeem~M. Mohideen}, \bibinfo{person}{Ping Hu},
  \bibinfo{person}{Ramesh Gupta}, \bibinfo{person}{Seeram Ramakrishna}, {and}
  \bibinfo{person}{Yong Liu}.} \bibinfo{year}{2021}\natexlab{}.
\newblock \showarticletitle{Realization of {{Circular Economy}} of {{3D Printed
  Plastics}}: A {{Review}}}.
\newblock \bibinfo{journal}{\emph{Polymers}} \bibinfo{volume}{13},
  \bibinfo{number}{5} (\bibinfo{year}{2021}), \bibinfo{pages}{744}.
\newblock


\bibitem[\protect\citeauthoryear{Zimmerman, Forlizzi, and Evenson}{Zimmerman
  et~al\mbox{.}}{2007}]%
        {zimmermanResearchDesignMethod2007}
\bibfield{author}{\bibinfo{person}{John Zimmerman}, \bibinfo{person}{Jodi
  Forlizzi}, {and} \bibinfo{person}{Shelley Evenson}.}
  \bibinfo{year}{2007}\natexlab{}.
\newblock \showarticletitle{Research through Design as a Method for Interaction
  Design Research in {{HCI}}}. In \bibinfo{booktitle}{\emph{Proceedings of the
  {{SIGCHI}} Conference on {{Human}} Factors in Computing Systems - {{CHI}}
  '07}}. \bibinfo{publisher}{{ACM Press}}, \bibinfo{address}{{San Jose,
  California, USA}}, \bibinfo{pages}{493}.
\newblock
\showISBNx{978-1-59593-593-9}
\urldef\tempurl%
\url{https://doi.org/10.1145/1240624.1240704}
\showDOI{\tempurl}


\end{thebibliography}

\end{document}